\documentclass[10pt, onecolumn]{article}

\usepackage[T1]{fontenc} 
\usepackage[utf8]{inputenc}
\usepackage{styles/arxiv}
\usepackage{listings}
\usepackage{subfiles}
\usepackage{syntax}
\usepackage{ebproof}
\usepackage{amssymb}
\usepackage{flagderiv}
\usepackage{newtxmath}

\DeclareUnicodeCharacter{22C4}{}

\title{Introducing Type Properties}
\author{Aziz Akhmedkhodjaev}

\begin{document}

\maketitle

\begin{abstract}
In type theory, we can express many practical ideas by
attributing some additional data to expressions we operate on during compilation.
For instance, some substructural type theories augment variables' typing judgments
with the information of their usage. That is, they allow one to explicitly state
how many times — \textit{0, 1, or many} — a variable can be used. This solves the problem
of resource usage control and allows us to treat variables as resources.

What's more, it often happens that this attributed information is interpreted (used)
during the same compilation and erased before we run a program.
A case in the point is that in the same substructural type theories,
their type checkers use these \textit{0, 1, or many}, to ensure that all variables
are used as many times as these attributions say them to be.

Yet, there wasn't any programming language concept whose concern would be to allow a programmer
to express these attributions in the language itself. That is, to let the programmer express
which data the one wants to attribute to what expressions and, most importantly, the meaning of the attributed data in their program.

As it turned out, the presence of such a concept allows us to express many practical
ideas in the language itself. For instance, with appropriate means for assigning
the meaning of these attributions, this concept would allow one to express linear types
as functionality in a separate program module, without the need to refine the whole
type system to add them.

In this paper, we present such a concept — we propose \textit{type properties}.
It allows a programmer to express these attributions while fulfilling the requirement
of being fully on the static level. That is, it allows one to express how to interpret
these attributions during compilation and erases them before a program is passed to the runtime.
\end{abstract}

\section{Introduction}

In type theory, when we want to perform static analysis or
record information about a program, we usually attribute
some additional data to expressions we operate on.
This can be illustrated by an approach called \textit{Quantitative Type Theory} (\cite{quantitative-type-theory}):

\[x_1 \stackrel{0}{:} S_1 ,\, x_2 \stackrel{1}{:} S_2 ,\, ...  \vdash M \stackrel{\sigma}{:} T\]

which augments its variables in typing contexts with \textit{0}, \textit{1}, and $\omega$, the elements of a semiring (\cite{semiring}).
In this system, the attributed data indicates how many times they are used at runtime:
\textit{0} indicates that a variable is not present at runtime, \textit{1} states that it is used only once,
while $\omega$ places no further restrictions and allows one to use the variable many times.
The explicit tracking of a variable's usage gives an opportunity to implement resource control usage (\cite{abel2018}).
After a type checker asserts that all variables were used as many times as they were said to be,
these attributed data are no longer needed, so it would be reasonable to erase them before
a program goes to the runtime.

As it turned out, resource usage control is not the only thing that
we can express by attributing additional data in this way.
To illustrate, in $\lambda_{Rust}$ (\cite{10.1145/3158154}), a reference's type is augmented with a lifetime,
which indicates that the reference is valid as long as the attributed lifetime is alive.
So, speaking about the expressive power of this technique, even static typing starts seeming
to be just a special case of this — its essence is to attribute types to terms,
perform the type check and erase types as a program is run
\footnote{
  What should be noted is that we speak about static typing in general
  without covering some specific cases. For instance, in dependent type
  theories, since types are treated as first-class values, sometimes one can
  propagate them to runtime just like any other data.
}.

The key \textbf{pattern} here is:
\begin{itemize}
      \item During compilation, we attribute some additional data to our expressions.
            A case in the point is those elements of the semiring — \textit{0}, \textit{1}, and $\omega$ —
            which are attributed to variables in the \textit{QTT}.
      \item During the same compilation, we use (interpret) these data.
            A case in the point is the assertion that all variables are used as many times
            as the attributions say them to be, again from the \textit{QTT}.
      \item We erase these attributed data since we already interpreted them
            during compilation so that they are not needed at runtime.
\end{itemize}

In the rest of this paper, data that are attributed to language expressions
like in the pattern above, will be called a \textit{static attribution}.

Yet, despite the shown expressive power of such static attributions,
there wasn't any programming language concept whose concern would be
the expression of them in the language itself. In other words, these static attributions
were kind of built-in in type theories, meaning that a programmer couldn't decide
what data the one wants to attribute to which language expressions, and, most importantly,
express what does the attributed data in their program mean.
The sentence above sounds quite abstract, so before we start a presentation
of a real type theory augmented with this concept, we explore its meaning in Section 2.

Back to the concept, it would allow us, for instance, to express that
we want to augment some variables with those \textit{0}, \textit{1}, and $\omega$, while the presence
of some means of "overloading" of the behavior of the case when a variable is
found to be used in an expression
\footnote{
  In other words, all we need is an opportunity to define such a function
  that is called when a variable of our (user-defined) type is found to
  be used in another expression. We can compare this to the notion of
  \textit{operators overloading} in modern programming languages, where
  it is possible to define such a function that is called when an operator
  on an object of a user-defined type is invoked. This can be seen as a generalization
  of such operators overloading since the function must be invoked when a variable is
  found to be used in any expression, not only in ones with operators.
} allows us to express that these elements of the semiring indicate how many times the variable is
used at runtime. In other words, this gives an opportunity to express the same linear types
just as functionality in a separate program module and include them whenever we need them.
That is, there is no need to refine the whole type theory to add them.
What should be noted is that it is the implementation of these static attributions that makes
the expression of linear types difficult in this way. It is not that hard to enlarge a theory with
such a notion of "overloading", whereas obliging a type theory to somehow remember what we attributed to what,
while fulfilling the requirement of being interpreted fully on the static level
(which is needed to, for instance, check the same linearity laws at compile-time) is a real challenge.

In this paper, we accept this challenge. We propose \textit{type properties} —
a programming language concept that allows a programmer to express these static attributions
the way we described — a user is capable of attributing any data to any expression and choose
the meaning of this attribution. The name of this concept comes from its implementation —
under the hood, it uses \textit{propertied types}, another notion introduced in this paper, to express them.

We also want to be clear and state explicitly what the reader should not expect from
this concept and this paper. The main purpose of this paper is to introduce this concept,
not to describe how it can be used to express something like linear types
\footnote{
  This is, in fact, the purpose of another upcoming research paper, on which we actively work now,
  and for which this paper, by introducing this concept, stands as a basis.
}.
What we also want to note is that the expression of something like linear types is obviously
not the only thing that this concept can help with. As said, the concept allows
a user to express static attributions themselves and leaves the question of their interpretation
on the programmer. That is, their meaning is limited only by other constructions of the language.

Our \textbf{contributions} are as follows:
\begin{itemize}
      \item Before we do the presentation of a real type theory augmented with this notion,
            we give its brief description and discuss how it can be potentially implemented (Section 2)
      \item Afterwards, we present a simple type theory, namely, $\lambda_{\twoheadrightarrow_p}$, where we formalize type properties (Section 3).
            Also, as we said many times, \textit{static} attributions must be handled during compilation
            and be absent at runtime, so, in the same section, we introduce an additional phase
            our programs must go through before they are run. We call this phase \textit{transformation},
            and it is the very phase where we handle static attributions.
      \item After the presentation, what we do is prove some important properties of the system (Section 4).
            In particular, we prove that promise that all static attributions are evaluated at compile-time,
            that the transformation is deterministic, and the property of the system of being type-sound.
      \item Finally, to make sure that it is clear how type properties are handled,
            we write two trivial programs and perform their step-by-step type checking,
            transformation, and runtime evaluation (Section 5).
\end{itemize}

What should be noted is that the absence of such a concept doesn't mean that
these static attributions weren't expressible in general — there have been type theories
that, in some sense, are capable of doing so.
These are theories where it is possible to implement other formal logic and
maybe even "embed" one in the system itself — in this case, the only thing one
needs to obtain, for instance, the same linear types, is just to formalize linear logic.
This, however, is not purely an expression of static attributions — it is just an expression
of a whole type theory that is augmented with ones.
But the real thing is that this fact, in any way, doesn't take away the need in such a concept —
in contrast, such theories may just formalize our notion to obtain the ability to express them.

\section{Type Properties}

First of all, let's imagine a type theory where it is possible
to express such static attributions and what does one must happen to have.
Let's also say that, for now, we are given that only data we can attribute to expressions
are ordinary integers.

The first thing that must be present is obviously the attribution itself:
\[
\begin{prooftree}
  \hypo{\Gamma \vdash 1 : \text{int}}
  \hypo{\Gamma \vdash n : \text{int}}
  \infer2[Set]{\Gamma \vdash 1^n : \text{int}}
\end{prooftree}
\]

As can be inferred from the rule above, by the superscript notation $M^n : \text{int}$ we denote
that \textit{M} is an expression of the type int, that has an attributed integer \textit{n}. 
When attributed, it is also necessary to have an ability to retrieve a static attribute back or even erase it:

\begin{equation*}
\begin{prooftree}
  \hypo{\Gamma \vdash 1^n : \text{int}}
  \infer1[Retrieve]{\Gamma \vdash n : \text{int}}
\end{prooftree}
\end{equation*}

\begin{equation*}
\begin{prooftree}
  \hypo{\Gamma \vdash 1^n : \text{int}}
  \infer1[Erase]{\Gamma \vdash 1 : \text{int}}
\end{prooftree}
\end{equation*}

Because an expression in this type theory is only optionally
attributed with an integer, it would be preferable to have an
opportunity to work with expressions, about which, during type check,
it is not yet known either they have an attributed integer or not.
For instance, if we have a function's argument — \textit{x} —
when type checking the body of the function, we don't know if an expression
that will be used to construct the value of the argument will have one.
All we need is a construction that would handle both cases — the presence and the absence of them.
Let's name this construction \textit{if-has}:

\[
\begin{prooftree}
  \hypo{\Gamma \vdash x^? : \text{int}}
  \hypo{\Gamma , x^n : \text{int}\, , n : \text{int} \vdash M : T}
  \hypo{\Gamma , x : \text{int} \vdash N : T}
  \hypo{(x^n,\, x,\,  n) \notin \Gamma}
  \infer4{\Gamma \vdash \text{if} \ x \ \text{has} \ n \ \text{then} \ M \ \text{else} \ N : T}
\end{prooftree}
\]

This construction cannot be evaluated during type checking, but at the same time, it can't be at runtime too
— we promised that all stuff related to static attributions is carried before we run our programs.
And that is exactly why our programs proceed through an additional phase before they are run —
the \textit{transformation} phase. At this phase, the construction \textit{if-has} can be evaluated easily.
This is because we already ensured that the body of the function is well-typed, so we can go further
and process its calls — and that's where we already know either the expression in the argument's position
has an attributed integer or not. Having this, we can process function monomorphization and propagate
the information about the attributed integer to the body of the function. When propagated,
the only thing \textit{if-has} will do is compile (yes, compile) itself to one of its branches,
according to the presence of the attribute.

Now, when we are equipped with the most basic operations on static attributions,
let's try to write a trivial program in pseudo-code and see what this program is transformed into.
Afterwards, we will finally discuss how we can use types to handle them.

One thing to mention — in a real programming language, most probably, you will never find
an expression with a superscript notation (like $1^n$), so, in our pseudo-code, we replaced this notation
with a more realistic one — set(\textit{1, n}).

\begin{lstlisting}
  greater_than_five (x : int) =
    if x has n then n > 5
    else x > 5
  
  let z = 1
  let z' = set(z, 1)

  let greater_z = greater_than_five z
  let greater_z' = greater_than_five z'
\end{lstlisting}

So, as it was pointed out in the introduction, a programmer is free to choose how to interpret these attributions.
In this trivial program, we agree that if there is a statically attributed constant to an expression
(such as $k$ in $x^k$), then it means that, during compilation, it is known that this expression will
be evaluated to the augmented constant.
With this in mind, we can implement a trivial optimization in our function \textit{greater_than_five} —
that is, by the construction \textit{if-has}, we can express that if there's an integer constant,
attributed to an expression that was used to construct the value of \textit{x}, then use it and perform
the comparison with respect to the constant, or to \textit{x}, otherwise.

According to the logic we described, our program will be transformed into something like:

\begin{lstlisting}
  greater_than_five_1 (x : int) = x > 5
  greater_than_five_2 (x : int) = 1 > 5
  
  let z = 1
  let z' = z

  let greater_z = greater_than_five_1 z
  let greater_z' = greater_than_five_2 z'
\end{lstlisting}

At this moment, we expressed a trivial optimization purely by the means of our theory.

Although it works, it's worth noting that our concept should never be used to express such optimizations in the code itself.
Any modern compiler can easily apply something like \textit{constant propagation} (\cite{10.1145/176454.176526}),
an optimization that computes constant expressions at compile-time rather than at runtime,
without any notion of type properties or type-theoretic static attributions and obtain even a better result.
Having that static attributions have much more useful applications,
we came up with this example because expressing something like from
the list in the introduction is a whole separate research topic.

Now let's finally see how we can use types to express static attributions and look at the real theory.

\paragraph{Using types to express static attributions}

First of all, let's make it clear what does it mean to "use types" to express these attributions.
Using types to express such attributions means that when a programmer wants to attribute one expression
to another (for instance, using the construction set(\textit{x, n}), as in the previous example), rather than attributing
to the term, just like we did before — $x^n : \text{int}$, we attribute it to its type. 

While, at first, this sounds weird, this approach has some serious advantages:

\begin{itemize}
  \item \textit{We don't augment type theory (in the sense of the study, not a type system) with something new and complex}.
        In the case when we attribute expressions to term themselves, then a new kind of terms,
        which is different from the ordinary one, arises.
        Having in mind that, in the real theory, we are not restricted to attributing only ordinary integers
        and the number of attributions is not fixed, we would need to extend our meta-theory to
        work with these terms too.
        In contrast, when we attribute those to types, what we get are types parametrized with terms —
        and this is what we are obviously familiar with — dependent types (\cite{intrDepIdr, WhyDepTypes, 10.1145/1411203.1411213})!
        Of course, they are not purely dependent types, since they are required to
        capture some other data too, but instead can be seen as a special case of ones or just a similar concept.
  \item By attributing expressions to types rather than to terms, we also likely to get the lifting of these attributions
        to the static level only. This is because most type theories (if we don't cover dependent cases)
        erase type data before a program goes to the runtime.
\end{itemize}

Now let's finally try to attribute something to a type rather than to term.
From now, let's also accept the fact that we can attribute constant expressions
of any type, not only integers, and the fact that the number of attributions
is not fixed (previously we could have only one attribution). According to our logic,
this would look like this:

\[
\begin{prooftree}
  \hypo{\Gamma \vdash 1 : \text{int}}
  \hypo{\Gamma \vdash n : \text{int}}
  \infer2{\Gamma \vdash \text{set}(1, eq, n) : [\text{int}] \langle eq \hookrightarrow n[\text{int}] \rangle}
\end{prooftree}
\]

First of all, because we can have arbitrarily many attributions, to later retrieve one of those,
we must somehow distinguish them among themselves — from now on, each attribute will come with a name.
Second, we have to use a new notation to perform attribution — we write set$(1, eq, n)$ to statically
attribute the expression $n$ to $1$, and give it the name $eq$.

Types that are constructed when one tries to perform an attribution in the rest of this paper will be called \textit{propertied types}.
Their notation — $[T] \langle p_1 \hookrightarrow e_1[P_1] , \cdots , p_n \hookrightarrow e_n[P_n] \rangle$ —
merely means that the original type of the term was $T$ and that one has static attributions in form of $p \hookrightarrow e[P]$,
where $p$, $e$, and $P$, indicate the name, expression, and type of the attribute, appropriately.
In this case, the type $T$ is said to be the base type of the propertied one.

But now we are in a trouble.
The thing is, our function \textit{greater_than_five}, expects an argument of the type int,
while we, from now, on line 9 in the previous example, pass the argument of a different type — of a propertied one, 
that is based on int. This means that the program won't even type check.

For this to work, we have to make our functions to be \textit{polymorphic with respect to type properties}.
This means that if a function accepts an argument of the type \textit{T}, then it also must accept an argument
of a propertied type that is based on \textit{T}.
This also means that to work as expected, we have to provide an efficient framework of function recompilation and monomorphization.
This is because if we place a construction like \textit{if-has} inside the body of a function, that will inspect the type
(since attributes are now kept in types) of the function's argument (as we did in \textit{greater_than_five}),
then the body of the function will depend on the type of the passed argument, which is not known since functions are polymorphic.
We can't go the simple way and do the inspection at runtime (\cite{10.1145/130697.130699}, \cite{10.1145/165180.165191})
since we promised that static attributions do not occur at one —
so, we are left only with one choice — to provide an efficient framework of function recompilation and monomorphization.

As the reader is already familiar with nearly what one has to expect from the rest of the chapters,
we think it is the very time to start the presentation of the real theory altogether with
things like the above-mentioned framework.

\section{The system $\lambda_{\twoheadrightarrow_p}$: Formalization of Type Properties}

In this section, we present a simple system augmented with the notion of type properties.
This system was originally based on \textit{Simply Typed Lambda Calculus},
but the augmentation of type properties has eliminated almost all observable similarities.
We call this system $\lambda_{\twoheadrightarrow p}$.

\subsection{Syntax}

The concrete syntax of $\lambda_{\twoheadrightarrow p}$ is defined as follows:

\begin{align*}
\begin{array}{lrllll}
    (\textit{typing contexts}) & \Gamma & ::= & ⋄ \ |\ \Gamma , x : T\\
    (\textit{types}) & T, P & ::= & \text{int} \ |\ \text{unit} \ |\  T_1 \rightarrow T_2\\
    (\textit{expressions}) & F, M, N, L, e, t & ::= & \text{func} \ f \ x : T \ \text{with} \ M \ \text{in} \ N \ |\  \text{let} \ x = e \ \text{in} \ L
    \\&&&|\  \text{if-has} \ L \ p : T \ \text{bind-as} \ x \ \text{in} \ M \ \text{else} \ N
    \\&&&|\  \text{extract}(M) \ |\  \text{set}(M, p, e) \ |\  \text{get}(M, p)
    \\&&&|\  \text{erase}(M, p) \ |\  M \ N \ |\  e_1 + e_2 \ |\  e_1 - e_2
    \\&&&|\  x \ |\ n\\
    (\textit{identifiers and variables}) & x, y, p\\
    (\textit{integer constants}) & n, k\\
\end{array}
\end{align*}

The system has two base types — \textit{int} and \textit{unit},
one compound — the function type, and a family of types that is not present in the syntax — the family of \textit{propertied types}.

In contrast with \textit{STLC}, this system does not have lambdas.
Instead, it has the construction \textit{func} that introduces a function in the scope of its continuation,
where one is able to refer to the former with a specific name.
The reason for having this will be made clear when we will start exploring how the system processes type properties.
The expressions \textit{set} and \textit{get} are responsible for setting and retrieving properties from and to types.
The expression \textit{if-has} checks if a type has a property of a specific type and binds
the value of the one to the identifier that follows the keyword \textit{bind-as},
and executes the corresponding code.
Finally, the constructions \textit{erase} and \textit{extract}
erase certain property of a type and extract the underlying value, the one that was given during formation, appropriately.

\subsection{Typing Rules}

Type formation judgment rules are as follows:

\[
\begin{prooftree}
    \hypo{}
    \infer1[F-Unit]{\Gamma \vdash \text{unit}}
\end{prooftree}
\]

\[
\begin{prooftree}
    \hypo{}
    \infer1[F-Int]{\Gamma \vdash \text{int}}
\end{prooftree}
\]

\[
\begin{prooftree}
    \hypo{\Gamma \vdash T_1}
    \hypo{\Gamma \vdash T_2}
    \infer2[F-Func]{\Gamma \vdash T_1 \rightarrow T_2}
\end{prooftree}
\]

\[
\begin{prooftree}
    \hypo{\Gamma \vdash T_1}
    \infer1[F-Prop-1]{\Gamma \vdash [T_1] \langle \rangle}
\end{prooftree}
\]

\[
\begin{prooftree}
    \hypo{\Gamma \vdash [T] \langle p_1 \hookrightarrow e_1[P_1] , \cdots , p_n \hookrightarrow e_n[P_n] \rangle}
    \hypo{\Gamma \vdash e : P}
    \hypo{( p_1 \neq p , \cdots , p_n \neq p )}
    \infer3[F-Prop-2]{
        \Gamma \vdash [T] \langle p_1 \hookrightarrow e_1[P_1] , \cdots , p_n \hookrightarrow e_n[P_n], p \hookrightarrow e[P] \rangle
    }
\end{prooftree}
\]

Here, $[T] \langle p_1 \hookrightarrow e_1[P_1] , \cdots , p_n \hookrightarrow e_n[P_n] \rangle$
represents a propertied type, which is based on the type \textit{T} and has properties
in form of $p \hookrightarrow e[P]$, where $p$ is an identifier that is used to refer to the property,
and $e$ is the corresponding expression of the property with the type $P$.
The notation $(p_1 \neq p , \cdots , p_n \neq p)$ in the rule \textit{F-Prop-2}
denotes that $p$ must not be already present in the properties.
We also implicitly assume that every rule in this section operates only on well-formed contexts.

The introduction rules go next:

\[
\begin{prooftree}
    \hypo{}
    \infer1[I-Unit]{ \vdash () : \text{unit} }
\end{prooftree}
\]

\[
\begin{prooftree}
    \hypo{}
    \infer1[I-Int]{ \vdash n : \text{int} }
\end{prooftree}
\]

\[
\begin{prooftree}
    \hypo{\Gamma \uplus \{ x : T_1 \} \vdash M : T_2}
    \infer[no rule]1{\Gamma \uplus \{ x : [T_1] \langle \rangle \} \vdash M : T_2}
    \hypo{\Gamma , f : T_1 \rightarrow T_2 \vdash e : T_e}
    \infer[no rule]1{(T_2 \neq [T] \langle \cdots \rangle) \wedge (T_2 \neq P_1 \rightarrow P_2)}
    \infer2[I-Func]{\Gamma \vdash (\text{func} \ f \ x : T_1 \ \text{with} \ M \ \text{in} \ e) : T_e}
\end{prooftree}
\]

\[
\begin{prooftree}
    \hypo{\Gamma \vdash M : T_1}
    \hypo{\Gamma \vdash e : T_2}
    \infer2[I-Prop-1]{
        \Gamma \vdash \text{set}(M, p, e) : [T_1] \langle p \hookrightarrow e[T_2] \rangle
    }
\end{prooftree}
\]

\[
\begin{prooftree}
    \hypo{\Gamma \vdash M : [T] \langle p_1 \hookrightarrow e_1[P_1] , \cdots , p_n \hookrightarrow e_n[P_n] \rangle}
    \hypo{\Gamma \vdash e : P}
    \hypo{( p_1 \neq p , \cdots , p_n \neq p )}
    \infer3[I-Prop-2]{
        \Gamma \vdash \text{set}(M, p, e) : [T] \langle p_1 \hookrightarrow e_1[P_1] , \cdots , p_n \hookrightarrow e_n[P_n], p \hookrightarrow e[P] \rangle
    }
\end{prooftree}
\]

\[
\begin{prooftree}
    \hypo{\Gamma \vdash M : [T] \langle props_1 , p \hookrightarrow e[P], props_2 \rangle}
    \hypo{\Gamma \vdash e' : P'}
    \infer2[I-Prop-3]{
        \Gamma \vdash \text{set}(M, p, e) : [T] \langle props_1 , p \hookrightarrow e'[P'], props_2 \rangle
    }
\end{prooftree}
\]

The notation $(T_2 \neq [T] \langle \cdots \rangle)$ indicates that the return type,
namely, $T_2$, cannot be a propertied type. In other words, we cannot return a value of a propertied type from a function.
Notice that we can't make our function accept values of ones too — there is no conventional syntax for propertied types
so that they exist only in our derivation rules.
The notation $\Gamma \uplus \{ x : T_1 \}$ means that we're adding the judgment $x : T_1$ to the context $\Gamma$
if $x \notin \Gamma$, and shadowing old with the new one otherwise.
The rule \textit{I-Prop-3} states that we can update the expression of a specific property in a propertied type.
In this rule, we wrote $props_1$ for all properties $p_1 \hookrightarrow e_1[P_1] , \cdots , p_n \hookrightarrow e_n[P_n]$
that come before the property with the name $p$, and $props_2$ for all that come after.

Rules that cover the rest of the expressions are present below.

\[
\begin{prooftree}
    \hypo{\Gamma \vdash t_1 : T_1 \rightarrow T_2}
    \hypo{\Gamma \vdash t_2 : T_1}
    \infer2[E-App-1]{\Gamma \vdash t_1 \ t_2 : T_2}
\end{prooftree}
\]

\[
\begin{prooftree}
    \hypo{\Gamma \vdash t_1 : T_1 \rightarrow T_2}
    \hypo{\Gamma \vdash t_2 : [T_1] \langle \cdots \rangle}
    \infer2[E-App-2]{\Gamma \vdash t_1 \ t_2 : T_2}
\end{prooftree}
\]

\[
\begin{prooftree}
    \hypo{\Gamma , x : T_1 \vdash M : T_2}
    \hypo{\Gamma \vdash N : T_1}
    \hypo{(T_2 \neq [T] \langle \cdots \rangle) \wedge (T_2 \neq P_1 \rightarrow P_2)}
    \infer3[E-Let]{\Gamma \vdash \text{let} \ x = N \ \text{in} \ M : T_2}
\end{prooftree}
\]

\[
\begin{prooftree}
    \hypo{\Gamma \vdash e_1 : \text{int}}
    \hypo{\Gamma \vdash e_2 : \text{int}}
    \infer2[E-Plus]{\Gamma \vdash e_1 + e_2 : \text{int}}
\end{prooftree}
\]

\[
\begin{prooftree}
    \hypo{\Gamma \vdash e_1 : [\text{int}] \langle \cdots \rangle}
    \hypo{\Gamma \vdash e_2 : \text{int}}
    \infer2[E-P-Plus-1]{\Gamma \vdash e_1 + e_2 : \text{int}}
\end{prooftree}
\]

\[
\begin{prooftree}
    \hypo{\Gamma \vdash e_1 : \text{int}}
    \hypo{\Gamma \vdash e_2 : [\text{int}] \langle \cdots \rangle}
    \infer2[E-P-Plus-2]{\Gamma \vdash e_1 + e_2 : \text{int}}
\end{prooftree}
\]

\[
\begin{prooftree}
    \hypo{\Gamma \vdash e_1 : [\text{int}] \langle \cdots \rangle}
    \hypo{\Gamma \vdash e_2 : [\text{int}] \langle \cdots \rangle}
    \infer2[E-P-Plus-3]{\Gamma \vdash e_1 + e_2 : \text{int}}
\end{prooftree}
\]

\[
\begin{prooftree}
    \hypo{\Gamma \vdash e_1 : \text{int}}
    \hypo{\Gamma \vdash e_2 : \text{int}}
    \infer2[E-Minus]{\Gamma \vdash e_1 - e_2 : \text{int}}
\end{prooftree}
\]

\[
\begin{prooftree}
    \hypo{\Gamma \vdash e_1 : [\text{int}] \langle \cdots \rangle}
    \hypo{\Gamma \vdash e_2 : \text{int}}
    \infer2[E-P-Minus-1]{\Gamma \vdash e_1 - e_2 : \text{int}}
\end{prooftree}
\]

\[
\begin{prooftree}
    \hypo{\Gamma \vdash e_1 : \text{int}}
    \hypo{\Gamma \vdash e_2 : [\text{int}] \langle \cdots \rangle}
    \infer2[E-P-Minus-2]{\Gamma \vdash e_1 - e_2 : \text{int}}
\end{prooftree}
\]

\[
\begin{prooftree}
    \hypo{\Gamma \vdash e_1 : [\text{int}] \langle \cdots \rangle}
    \hypo{\Gamma \vdash e_2 : [\text{int}] \langle \cdots \rangle}
    \infer2[E-P-Minus-3]{\Gamma \vdash e_1 - e_2 : \text{int}}
\end{prooftree}
\]

\[
\begin{prooftree}
    \hypo{\Gamma \vdash M : [T] \langle \cdots , p \hookrightarrow e[P] , \cdots \rangle}
    \infer1[E-Get-Prop]{\Gamma \vdash \text{get}(M, p) : P}
\end{prooftree}
\]

\[
\begin{prooftree}
    \hypo{\Gamma , L : T_L \vdash T_x}
    \hypo{\Gamma , L : [T_L] \langle \rangle \vdash N : T}
    \infer[no rule]2{\Gamma \uplus \{ x : T_x \} , L : [T_L] \langle p \hookrightarrow x[T_x] \rangle \vdash M : T}
    \hypo{(e \notin \Gamma) \wedge (T_L \neq [T'] \langle \rangle)}
    \infer2[E-If-Has-1]{\Gamma , L : T_L \vdash \text{if-has} \ L \ p : T_x \ \text{bind-as} \ x \ \text{in} \ M \ \text{else} \ N : T}
\end{prooftree}
\]

\[
\begin{prooftree}
    \hypo{\Gamma , L : [T_L] \langle props \rangle \vdash T_x}
    \hypo{\Gamma , L : [T_L] \langle props \smallsetminus p \rangle \vdash N : T}
    \infer[no rule]2{\Gamma \uplus \{ x : T_x \} , L : [T_L] \langle props \smallsetminus p , p \hookrightarrow x[T_x] \rangle \vdash M : T}
    \hypo{(e \notin \Gamma)}
    \infer2[E-If-Has-2]{\Gamma , L : [T_L] \langle props \rangle \vdash \text{if-has} \ L \ p : T_x \ \text{bind-as} \ x \ \text{in} \ M \ \text{else} \ N : T}
\end{prooftree}
\]

\[
\begin{prooftree}
    \hypo{\Gamma \vdash M : [T] \langle \cdots \rangle}
    \infer1[E-Ext]{\Gamma \vdash \text{extract}(M) : T}
\end{prooftree}
\]

\[
\begin{prooftree}
    \hypo{\Gamma \vdash M : [T] \langle props_1 , p \hookrightarrow e[P] , props_2 \rangle}
    \infer1[E-Erase]{\Gamma \vdash \text{erase}(M, p) : [T] \langle props_1 , props_2 \rangle}
\end{prooftree}
\]

One may notice that the rule \textit{E-Let} prohibits its continuation to be of a propertied or function type.
The reason for that will be made clear when we will explore how the system processes type properties.
The rule \textit{E-App-2} is one of the main rules that make functions be polymorphic with respect to propertied types.
Namely, it states that if a function accepts an argument of the type \textit{$T_1$}, then it also accepts an argument
of a propertied type that is based on \textit{$T_1$}. Another interesting rule is the rule \textit{E-If-Has-1} —
it states that if one tries to check if a property is assigned on even a non-propertied type, then the expression is still well-typed.
This is because when one checks if the body of a function is well-typed and tries to check if a type
has an assigned property, the type of the argument itself may not be a propertied one, but instead the base type of the one.

Finally, the structural rules are:

\[
\begin{prooftree}
    \infer0[S-Var]{\Gamma , x : T , \Gamma' \vdash x : T}
\end{prooftree}
\]

\[
\begin{prooftree}
    \hypo{\Gamma \vdash L : T_L}
    \hypo{\Gamma \vdash T_x}
    \hypo{(x \notin \Gamma)}
    \infer3[S-Weak]{\Gamma , x : T_x \vdash L : T_L}
\end{prooftree}
\]

\[
\begin{prooftree}
    \hypo{\Gamma , \Gamma' \vdash L : T_L}
    \infer1[S-Exchange]{\Gamma' , \Gamma \vdash L : T_L}
\end{prooftree}
\]

\subsection{Processing type properties}

As said in Section 2, type properties must live only in compile-time.
This means that we must somehow evaluate them, and this evaluation must be processed right before a program is passed to the runtime.
For this purpose, we introduce a new judgment — the judgment $\twoheadrightarrow_p$,
which is a big-step operational semantics judgment that only evaluates stuff with type properties.
In the rest of this paper, this form of evaluation will be called \textit{transformation}.

An efficient implementation of type properties would utilize effective means of function recompilation.
In order to perform any recompilation stuff, we must somehow save bodies and information of functions defined,
so we introduce a context that serves the desired purpose.
We call this context \textit{functional context} $\Delta$ and define it as follows:

\[\Delta ::= ⋄ \ | \ \Delta , f :: x : T_1 \, . \, M : T_2 \ | \ \Delta , f \, [n] \  \triangleright x : T_1 \, . \, M : T_2 \]

Since functions in our system are not anonymous,
to associate something with a specific function,
it is enough to associate it with its name.
If our language would instead have \textit{ordinary lambdas} (or any other type of anonymous functions),
other facilities for body propagation, which would, by chance, be more complicated than just named functions, must be utilized.

$f :: x : T_1 \, . \, M : T_2$ indicates that $f$ is a function
that takes $x$ of the type $T_1$ as an argument, its body is $M$, and the resulting type is $T_2$.
It stands for a function that contains a raw code (not transformed yet) and therefore cannot be propagated to a runtime call.
$f \, [n] \  \triangleright x : T_1 \, . \, M : T_2$, where $n$ is any natural number,
stands for a monomorphized version of \textit{f}, meaning that its body
has already gone through transformation and it is ready to be called.
$n$ here stands just for distinguishing different monomorphizations of the function \textit{f}.

Below we list some notations that we'll use in this section

\begin{itemize}
    \item 
    $f \notin \Delta$ — there is no entry $f :: x : T_1 \, . \, M : T_2$,
    where $T_1$ and $T_2$ match \textit{arbitrary} types and $M$ matches \textit{any} term,
    in the context $\Delta$.
    
    \item
    $f \, [n] \notin \Delta$ — there is no entry $f \, [k] \triangleright x : T_1 \, . \, M : T_2$,
    where $T_1$ and $T_2$ match \textit{arbitrary} types, $M$ matches \textit{any} term,
    and $k$ matches the specific number $n$, in the context $\Delta$.

    \item
    $f \, [n] \triangleright x : T_1 \, . \, M : T_2 \notin \Delta$ — there is no entry exactly
    this entry in the context $\Delta$.
\end{itemize}

Functional contexts are processed by the new judgment,
while the latter itself works under a typing one:

\[\Gamma ::= ⋄ \ | \ \Gamma , x : T \ | \ \Gamma , \langle \Delta_1 ; t_1 \rangle \twoheadrightarrow_p \langle \Delta_2 ; t_2 \rangle : T \]

To illustrate, the judgment $\Gamma \vdash \langle \Delta_1 ; t_1 \rangle \twoheadrightarrow_p \langle \Delta_2 ; t_2 \rangle : T$
merely means that the program $t_1$, under the functional context $\Delta$,
is transformed into the program $t_2$, the functional context became $\Delta'$,
and the type of $t_2$ is $T$.

We also extend structural rules that work on typing context to cover the transformation judgment as follows:

\[
\begin{prooftree}
    \infer0[R-S-Red]{
        \Gamma , \langle \Delta ; L \rangle \twoheadrightarrow_p \langle \Delta' ; L' \rangle : T , \Gamma'
        \vdash \langle \Delta ; L \rangle \twoheadrightarrow_p \langle \Delta' ; L' \rangle : T
    }
\end{prooftree}
\]

\[
\begin{prooftree}
    \hypo{\Gamma \vdash M : T_M}
    \hypo{\Gamma \vdash T_L}
    \hypo{\Gamma \vdash L : T_L}
    \hypo{\Gamma \vdash L' : T_L}
    \hypo{(\langle \Delta ; L \rangle \twoheadrightarrow_p \langle \Delta' ; L' \rangle \notin \Gamma)}
    \infer5[R-S-Weak-1]{\Gamma, \langle \Delta ; L \rangle \twoheadrightarrow_p \langle \Delta' ; L' \rangle : T_L \vdash M : T_M }
\end{prooftree}
\]

\[
\begin{prooftree}
    \hypo{(\langle \Delta_L ; L \rangle \twoheadrightarrow_p \langle \Delta_L' ; L' \rangle \notin \Gamma)}
    \infer[no rule]1{\Gamma \vdash \langle \Delta_M ; M \rangle \twoheadrightarrow_p \langle \Delta_M' ; M' \rangle : T_M}
    \hypo{\Gamma \vdash L : T_L}
    \infer[no rule]1{\Gamma \vdash L' : T_L}
    \hypo{\Gamma \vdash T_L}
    \infer3[R-S-Weak-2]{
        \Gamma , \langle \Delta_L ; L \rangle \twoheadrightarrow_p \langle \Delta_L' ; L' \rangle : T_L
        \vdash \langle \Delta_M ; M \rangle \twoheadrightarrow_p \langle \Delta_M' ; M' \rangle : T_M
    }
\end{prooftree}
\]

\[
\begin{prooftree}
    \hypo{\Gamma \vdash \langle \Delta ; L \rangle \twoheadrightarrow_p \langle \Delta' ; L' \rangle : T_L}
    \hypo{\Gamma \vdash T_x}
    \hypo{(x \notin \Gamma)}
    \infer3[R-S-Weak-3]{\Gamma , x : T_x \vdash \langle \Delta ; L \rangle \twoheadrightarrow_p \langle \Delta' ; L' \rangle : T_L }
\end{prooftree}
\]

\[
\begin{prooftree}
    \hypo{\Gamma , \Gamma' \vdash \langle \Delta ; L \rangle \twoheadrightarrow_p \langle \Delta' ; L' \rangle : T_L}
    \infer1[R-S-Exchange]{\Gamma' , \Gamma \vdash \langle \Delta ; L \rangle \twoheadrightarrow_p \langle \Delta' ; L' \rangle : T_L}
\end{prooftree}
\]

\[
\begin{prooftree}
    \infer0[R-S-Var]{\Gamma , \langle \Delta ; L \rangle \twoheadrightarrow_p \langle \Delta' ; L' \rangle : T , \Gamma' \vdash L' : T}
\end{prooftree}
\]

\[
\begin{prooftree}
    \hypo{(f \notin \Delta \cup \Delta')}
    \infer[no rule]1{\Gamma \vdash \langle \Delta ; L \rangle \twoheadrightarrow_p \langle \Delta' ; L' \rangle : T_L}
    \hypo{\Gamma , x : T_x \vdash M : T_M}
    \infer[no rule]1{\Gamma , x : [T_x] \langle \rangle \vdash M : T_M}
    \infer2[R-S-Weak-Delta-1]{
        \Gamma \vdash
        \langle \Delta , f :: x : T_x \ . \ M : T_M ; L \rangle
        \twoheadrightarrow_p
        \langle \Delta' , f :: x : T_x \ . \ M : T_M ; L' \rangle : T_L
    }
\end{prooftree}
\]

\[
\begin{prooftree}
    \hypo{(f \, [n] \notin \Delta \cup \Delta')}
    \infer[no rule]1{\Gamma \vdash \langle \Delta ; L \rangle \twoheadrightarrow_p \langle \Delta' ; L' \rangle : T_L}
    \hypo{\Gamma , x : T_x \vdash M : T_M}
    \infer[no rule]1{\Gamma , x : [T_x] \langle \rangle \vdash M : T_M}
    \infer2[R-S-Weak-Delta-2]{
        \Gamma \vdash
        \langle \Delta , f \, [n] \triangleright x : T_x \ . \ M : T_M ; L \rangle
        \twoheadrightarrow_p
        \langle \Delta' , f \, [n] \triangleright x : T_x \ . \ M : T_M ; L' \rangle : T_L
    }
\end{prooftree}
\]

\[
\begin{prooftree}
    \hypo{\Gamma \vdash \langle \Delta_1 , \Delta_2 ; L \rangle \twoheadrightarrow_p \langle \Delta_1' , \Delta_2' ; L' \rangle : T_L}
    \infer1[R-S-Exchange-Delta]{\Gamma \vdash \langle \Delta_2 , \Delta_1 ; L \rangle \twoheadrightarrow_p \langle \Delta_1' , \Delta_2' ; L' \rangle : T_L}
\end{prooftree}
\]

Constants and literals have nothing to do with type properties, so they are transformed into themselves:

\[
\begin{prooftree}
    \hypo{}
    \infer1[R-V-Unit]{\vdash \, \langle ; () \rangle \twoheadrightarrow_p \, \langle ; () \rangle : \text{unit}}
\end{prooftree}
\]

\[
\begin{prooftree}
    \hypo{}
    \infer1[R-V-Int]{\vdash \, \langle ; n \rangle \twoheadrightarrow_p \, \langle ; n \rangle : \text{int}}
\end{prooftree}
\]

\[
\begin{prooftree}
    \hypo{}
    \infer1[R-V-Func]{\vdash \langle f :: x : T_1 \ . \ M : T_2 ; f \rangle \twoheadrightarrow_p \langle f :: x : T_1 \ . \ M : T_2 ; f \rangle : T_1 \rightarrow T_2}
\end{prooftree}
\]

Now we present the rules that work with type properties directly so that they somehow transform the expressions they operate on.
Rules for the expressions that are used to set properties go first.

\[
\begin{prooftree}
    \hypo{\Gamma \vdash \langle \Delta ; M \rangle \twoheadrightarrow_p \langle \Delta_M ; M' \rangle : T}
    \hypo{\Gamma \vdash \langle \Delta_M ; e \rangle \twoheadrightarrow_p \langle \Delta' ; e' \rangle : P }
    \hypo{(T \neq [T'] \langle \cdots \rangle)}
    \infer3[R-Set-1]{
        \Gamma \vdash
        \langle \Delta ; \text{set}(M, p, e) \rangle \twoheadrightarrow_p \langle \Delta' ; \text{propertied}[M'] \rangle : [T] \langle p \hookrightarrow e'[P] \rangle
    }
\end{prooftree}
\]

\[
\begin{prooftree}
    \hypo{\Gamma \vdash \langle \Delta_M ; e \rangle \twoheadrightarrow_p \langle \Delta' ; e' \rangle : P }
    \hypo{( p_1 \neq p , \cdots , p_n \neq p )}
    \infer[no rule]2{
        \Gamma \vdash \langle \Delta ; M \rangle \twoheadrightarrow_p \langle \Delta_M ; \text{propertied}[M'] \rangle
            : [T] \langle p_1 \hookrightarrow e_1[P_1] , \cdots , p_n \hookrightarrow e_n[P_n] \rangle
    }
    \infer1[R-Set-2]{
        \Gamma \vdash \langle \Delta ; \text{set}(M, p, e) \rangle \twoheadrightarrow_p \langle \Delta' ; \text{propertied}[M'] \rangle
            : [T] \langle p_1 \hookrightarrow e_1[P_1] , \cdots , p_n \hookrightarrow e_n[P_n], p \hookrightarrow e[P] \rangle
    }
\end{prooftree}
\]

\[
\begin{prooftree}
    \hypo{\Gamma \vdash \langle \Delta ; M \rangle \twoheadrightarrow_p \langle \Delta_M ; \text{propertied}[M'] \rangle
        : [T] \langle props_1 , p \hookrightarrow e[P], props_2 \rangle}
    \hypo{\Gamma \vdash \langle \Delta_M ; e \rangle \twoheadrightarrow_p \langle \Delta' ; e' \rangle : P }
    \infer2[R-Set-3]{
        \Gamma \vdash \langle \Delta ; \text{set}(M, p, e) \rangle \twoheadrightarrow_p \langle \Delta' ; \text{propertied}[M'] \rangle
            : [T] \langle props_1 , p \hookrightarrow e[P'] , props_2 \rangle
    }
\end{prooftree}
\]

The rules above all transform their expressions to $\text{propertied}[\cdots]$,
which we will use to intermediately (during compile-time program transformation)
represent any value of a propertied type, with the underlying value \textit{x}.
For instance, if we would set a property \textit{equal\_to} with the value of \textit{5} to the integer constant \textit{5},
then we would get the term $\text{propertied}[5] : [\text{int}] \langle equal\_to \hookrightarrow 5[int] \rangle$.

Next go the rules for retrieving and erasing type properties, as well as extracting underlying values:

\[
\begin{prooftree}
    \hypo{
        \Gamma \vdash \langle \Delta ; M \rangle \twoheadrightarrow_p \langle \Delta_M ; M' \rangle : [T] \langle \cdots , p \hookrightarrow e[P] , \cdots \rangle
    }
    \hypo{\Gamma \vdash \langle \Delta_M ; e \rangle \twoheadrightarrow_p \langle \Delta_e ; e' \rangle : P}
    \infer2[R-Get]{\Gamma \vdash \langle \Delta ; \text{get}(M, p) \rangle \twoheadrightarrow_p \langle \Delta_e ; e' \rangle : P}
\end{prooftree}
\]

\[
\begin{prooftree}
    \hypo{\Gamma \vdash \langle \Delta ; M \rangle \twoheadrightarrow_p \langle \Delta' ; \text{propertied}[M'] \rangle : [T] \langle \cdots \rangle}
    \infer1[R-Ext]{\Gamma \vdash \langle \Delta ; \text{extract}(M) \rangle \twoheadrightarrow_p \langle \Delta' ; M' \rangle : T}
\end{prooftree}
\]

\[
\begin{prooftree}
    \hypo{
        \Gamma \vdash
        \langle \Delta ; M \rangle
        \twoheadrightarrow_p
        \langle \Delta' ; \text{propertied}[M'] \rangle : [T] \langle props_1 , p \hookrightarrow e[P] , props_2 \rangle
    }
    \infer1[R-Erase]{
        \Gamma \vdash
        \langle \Delta ; \text{erase}(M, p) \rangle
        \twoheadrightarrow_p
        \langle \Delta' ; \text{propertied}[M'] \rangle : [T] \langle props_1 , props_2 \rangle
    }
\end{prooftree}
\]

Notice that the expression \textit{get}, in contrast to \textit{if-has} works only on expressions of propertied types.
This means that we are unable to retrieve a property from a function's argument without checking that it is present,
since it is impossible for a function to work only on propertied types — they must work on their base types too.

\[
\begin{prooftree}
    \hypo{
        (T_L \neq [T'] \langle \cdots \rangle) \wedge
        (E \notin \Gamma) \wedge
        (\langle \Delta_1 ; E \rangle \twoheadrightarrow_p \langle \Delta_2 ; E' \rangle : T_E \notin \Gamma)
    }
    \infer[no rule]1{\Gamma \vdash \langle \Delta ; L \rangle \twoheadrightarrow_p \langle \Delta_L ; L' \rangle : T_L}
    \infer[no rule]1{
        \Gamma ,
        \langle \Delta ; E \rangle
        \twoheadrightarrow_p
        \langle \Delta_L ; \text{propertied}[L'] \rangle : [T_L] \langle \rangle \vdash \langle \Delta_L ; N[L/E] \rangle \twoheadrightarrow_p \langle \Delta_N ; N' \rangle : T_N
    }
    \infer1[R-If-Has-1]{
        \Gamma \vdash \langle \Delta ; \text{if-has} \ L \ p : T_x \ \text{bind-as} \ x \ \text{in} \ M \ \text{else} \ N \rangle \twoheadrightarrow_p \langle \Delta_N ; N' \rangle : T_N
    }
\end{prooftree}
\]

\[
\begin{prooftree}
    \hypo{\Gamma \vdash \langle \Delta_L ; N \rangle \twoheadrightarrow_p \langle \Delta_N ; N' \rangle : T_N}
    \hypo{( p_1 \neq p , \cdots , p_n \neq p )}
    \infer[no rule]2{
        \Gamma \vdash \langle \Delta ; L \rangle \twoheadrightarrow_p \langle \Delta_L ; L' \rangle : [T] \langle p_1 \hookrightarrow e_1[P_1] , \cdots , p_n \hookrightarrow e_n[P_n] \rangle
    }
    \infer1[R-If-Has-2]{
        \Gamma \vdash \langle \Delta ; \text{if-has} \ L \ p : T_x \ \text{bind-as} \ x \ \text{in} \ M \ \text{else} \ N \rangle \twoheadrightarrow_p \langle \Delta_N ; N' \rangle : T_N
    }
\end{prooftree}
\]

\[
\begin{prooftree}
    \hypo{\Gamma \vdash \langle \Delta_L ; N \rangle \twoheadrightarrow_p \langle \Delta_N ; N' \rangle : T_N}
    \hypo{(P \neq T_x)}
    \infer[no rule]2{\Gamma \vdash \langle \Delta ; L \rangle \twoheadrightarrow_p \langle \Delta_L ; L' \rangle : [T] \langle \cdots , p \hookrightarrow e[P] , \cdots \rangle}
    \infer1[R-If-Has-3]{
        \Gamma \vdash \langle \Delta ; \text{if-has} \ L \ p : T_x \ \text{bind-as} \ x \ \text{in} \ M \ \text{else} \ N \rangle \twoheadrightarrow_p \langle \Delta_N ; N' \rangle : T_N
    }
\end{prooftree}
\]

\[
\begin{prooftree}
    \hypo{\Gamma \vdash \langle \Delta_L ; e \rangle \twoheadrightarrow_p \langle \Delta_e ; e' \rangle : P}
    \infer[no rule]1{\Gamma \vdash \langle \Delta ; L \rangle \twoheadrightarrow_p \langle \Delta_L ; L' \rangle : [T] \langle \cdots , p \hookrightarrow e[P] , \cdots \rangle}
    \infer[no rule]1{
        \Gamma \uplus \{ \langle \Delta_L ; x \rangle \twoheadrightarrow_p \langle \Delta_e ; x \rangle : P \}
            \vdash \langle \Delta_e ; M \rangle \twoheadrightarrow_p \langle \Delta_M ; M' \rangle : T_M
    }
    \infer[no rule]1{(P \neq T_1 \rightarrow T_2) \wedge (T_M \neq T_1 \rightarrow T_2) \wedge (T_M \neq [T_P] \langle \cdots \rangle)}
    \infer1[R-If-Has-4]{
        \Gamma \vdash
        \langle \Delta ; \text{if-has} \ L \ p : P \ \text{bind-as} \ x \ \text{in} \ M \ \text{else} \ N \rangle
        \twoheadrightarrow_p
        \langle \Delta_M ; \text{let} \ x = e'\ \text{in} \ M' \rangle : T_M
    }
\end{prooftree}
\]

\[
\begin{prooftree}
    \hypo{\Gamma \vdash \langle \Delta_L ; e \rangle \twoheadrightarrow_p \langle \Delta_e ; e' \rangle : P}
    \infer[no rule]1{\Gamma \vdash \langle \Delta ; L \rangle \twoheadrightarrow_p \langle \Delta_L ; L' \rangle : [T] \langle \cdots , p \hookrightarrow e[P] , \cdots \rangle}
    \infer[no rule]1{
        \Gamma \uplus \{ \langle \Delta_L ; x \rangle \twoheadrightarrow_p \langle \Delta_e ; \, e' \rangle : P \}
        \vdash \langle \Delta_e ; M \rangle \twoheadrightarrow_p \langle \Delta_M ; M' \rangle : T_M
    }
    \infer[no rule]1{(P = T_1 \rightarrow T_2) \lor (T_M = T_1 \rightarrow T_2) \lor (T_M = [T_P] \langle \cdots \rangle)}
    \infer1[R-If-Has-5]{
        \Gamma \vdash
        \langle \Delta ; \text{if-has} \ L \ p : P \ \text{bind-as} \ x \ \text{in} \ M \ \text{else} \ N \rangle
        \twoheadrightarrow_p
        \langle \Delta_M ; M' \rangle : T_M
    }
\end{prooftree}
\]

The rules above define the transformation of the expression \textit{if-has}.
To work as expected, the value of $L$, whether it is of a propertied type or not,
must enter the branch \textit{else} with the same type.
Because of that, in the rule \textit{R-If-Has-1}, we made $L$ be transformed into the value of a propertied type
with the base type of \textit{$T_L$} and without any properties.
Then, $L$, whether from the rule \textit{R-If-Has-1}, \textit{R-If-Has-2} or \textit{R-If-Has-3},
always enters the branch \textit{else} being of a propertied type.

The transformation rule for a function definition expression is as follows:

\[
\begin{prooftree}
    \hypo{\Gamma \uplus \{ x : T_1 \} \vdash M : T_2}
    \hypo{\Gamma \uplus \{ f : T_1 \rightarrow T_2 \} \vdash \langle \Delta , f :: x : T_1 \ . \ M : T_2 ; e \rangle \twoheadrightarrow_p \langle \Delta' ; e' \rangle : T_e}
    \infer2[R-Func]{\Gamma \vdash \langle \Delta ; \text{func} \ f \ x : T_1 \ \text{with} \ M \ \text{in} \ e \rangle \twoheadrightarrow_p \langle \Delta' ; e' \rangle : T_e}
\end{prooftree}
\]

The requirement on the typing judgment $\Gamma \uplus \{ x : T_1 \} \vdash M : T_2$
is only needed to infer the type $T_2$, not to ensure that the body is well-typed.
This is due to the fact that before transformation, we expect the program to pass the type checker.
The rules below are the main rules that allow functions to be polymorphic with respect to type properties.

\[
\begin{prooftree}
    \hypo{(f \, [k] \notin \Delta \cup \Delta_x)}
    \infer[no rule]1{(\Delta_f \equiv f :: x : T_1 \ . \ M : T_2)}
    \infer[no rule]1{(\Delta_{f [k]} \equiv f \, [k] \, \triangleright x : T_1 \ . \ M' : T_2)}
    \infer[no rule]1{\Gamma \vdash \langle \Delta_i ; F \rangle \twoheadrightarrow_p \langle \Delta , \Delta_f ; f \rangle : T_1 \rightarrow T_2}
    \infer[no rule]1{
        \Gamma \vdash
        \langle \Delta , \Delta_f ; N \rangle
        \twoheadrightarrow_p
        \langle \Delta_x , \Delta_f ; \text{propertied}[y] \rangle : [T_1] \langle props \rangle
    }
    \infer[no rule]1{
        \Gamma \uplus \{ \langle \Delta ; x \rangle \twoheadrightarrow_p \langle \Delta_x ; \text{propertied}[x] \rangle: [T_1] \langle props \rangle \}
        \vdash \langle \Delta_x ; M \rangle \twoheadrightarrow_p \langle \Delta_M ; M' \rangle : T_2
    }
    \infer1[R-App-Compile-Prop-1]{
        \Gamma \vdash
        \langle \Delta_i ; F N \rangle
        \twoheadrightarrow_p
        \langle \Delta_M , \Delta_f, \Delta_{f [k]} ; f \, [k] \ y \rangle : T_2
    }
\end{prooftree}
\]

\[
\begin{prooftree}
    \hypo{(\Delta_f \equiv f :: x : T_1 \ . \ M : T_2)}
    \infer[no rule]1{(\Delta_{f [k]} \equiv f \, [k] \, \triangleright x : T_1 \ . \ M' : T_2)}
    \infer[no rule]1{\Gamma \vdash \langle \Delta_i ; F \rangle \twoheadrightarrow_p \langle \Delta , \Delta_f ; f \rangle : T_1 \rightarrow T_2}
    \infer[no rule]1{
        \Gamma \vdash
        \langle \Delta , \Delta_f; N \rangle
        \twoheadrightarrow_p
        \langle \Delta_x , \Delta_f , \Delta_{f [k]}; \text{propertied}[y] \rangle : [T_1] \langle props \rangle
    }
    \infer[no rule]1{
        \Gamma \uplus \{\langle \Delta ; x \rangle \twoheadrightarrow_p \langle \Delta_x ; \text{propertied}[x] \rangle : [T_1] \langle props \rangle \}
        \vdash \langle \Delta_x ; M \rangle \twoheadrightarrow_p \langle \Delta_M ; M' \rangle : T_2
    }
    \infer1[R-App-Ready-Prop-1]{
        \Gamma \vdash
        \langle \Delta_i ; F N \rangle
        \twoheadrightarrow_p
        \langle \Delta_M , \Delta_f , \Delta_{f [k]}; f \, [k] \ y \rangle : T_2
    }
\end{prooftree}
\]

\[
\begin{prooftree}
    \hypo{(f \, [k] \notin \Delta \cup \Delta_{f'})}
    \infer[no rule]1{(\Delta_f \equiv f :: x : P_1 \rightarrow P_2 \ . \ M : T_M)}
    \infer[no rule]1{(\Delta_{f [k]} \equiv f \, [k] \, \triangleright x : P_1 \rightarrow P_2 \ . \ M' : T_M)}
    \infer[no rule]1{\Gamma \vdash \langle \Delta_i ; F \rangle \twoheadrightarrow_p \langle \Delta , \Delta_f ; f \rangle : (P_1 \rightarrow P_2) \rightarrow T_M}
    \infer[no rule]1{
        \Gamma \vdash
        \langle \Delta , \Delta_f ; N \rangle
        \twoheadrightarrow_p
        \langle \Delta_{f'} , \Delta_f ; \text{propertied}[f'] \rangle
        : [P_1 \rightarrow P_2] \langle props \rangle
    }
    \infer[no rule]1{
        \Gamma \uplus \{\langle \Delta ; x \rangle \twoheadrightarrow_p \langle \Delta_{f'} ; \text{propertied}[f'] \rangle : [P_1 \rightarrow P_2] \langle props \rangle \}
        \vdash \langle \Delta_{f'} ; M \rangle \twoheadrightarrow_p \langle \Delta_M ; M' \rangle : T_M
    }
    \infer1[R-App-Compile-Prop-2]{
        \Gamma \vdash
        \langle \Delta_i ; F N \rangle
        \twoheadrightarrow_p
        \langle \Delta_M , \Delta_f, \Delta_{f [k]} ; f \, [k] \, f' \rangle : T_M
    }
\end{prooftree}
\]

\[
\begin{prooftree}
    \hypo{(\Delta_f \equiv f :: x : P_1 \rightarrow P_2 \ . \ M : T_M)}
    \infer[no rule]1{(\Delta_{f [k]} \equiv f \, [k] \, \triangleright x : P_1 \rightarrow P_2 \ . \ M' : T_M)}
    \infer[no rule]1{\Gamma \vdash \langle \Delta_i ; F \rangle \twoheadrightarrow_p \langle \Delta , \Delta_f ; f \rangle : (P_1 \rightarrow P_2) \rightarrow T_M}
    \infer[no rule]1{
        \Gamma \vdash
        \langle \Delta , \Delta_f; N \rangle
        \twoheadrightarrow_p
        \langle \Delta_{f'} , \Delta_f , \Delta_{f [k]}; \text{propertied}[f'] \rangle : [P_1 \rightarrow P_2] \langle props \rangle
    }
    \infer[no rule]1{
        \Gamma \uplus \{\langle \Delta ; x \rangle \twoheadrightarrow_p \langle \Delta_{f'} ; \text{propertied}[f'] \rangle : [P_1 \rightarrow P_2] \langle props \rangle \}
        \vdash \langle \Delta_{f'} ; M \rangle \twoheadrightarrow_p \langle \Delta_M ; M' \rangle : T_M}
    \infer1[R-App-Ready-Prop-2]{
        \Gamma \vdash
        \langle \Delta_i ; F N \rangle
        \twoheadrightarrow_p
        \langle \Delta_M , \Delta_f , \Delta_{f [k]}; f \, [k] \ f' \rangle : T_M
    }
\end{prooftree}
\]

\[
\begin{prooftree}
    \hypo{(f \, [k] \notin \Delta \cup \Delta_{f'})}
    \infer[no rule]1{(\Delta_f \equiv f :: x : P_1 \rightarrow P_2 \ . \ M : T_M)}
    \infer[no rule]1{(\Delta_{f [k]} \equiv f \, [k] \, \triangleright x : P_1 \rightarrow P_2 \ . \ M' : T_M)}
    \infer[no rule]1{
        \Gamma \vdash
        \langle \Delta , \Delta_f; N \rangle
        \twoheadrightarrow_p
        \langle \Delta_{f'} , \Delta_f; f' \rangle : P_1 \rightarrow P_2
    }
    \infer[no rule]1{\Gamma \vdash \langle \Delta_i ; F \rangle \twoheadrightarrow_p \langle \Delta , \Delta_f ; f \rangle : (P_1 \rightarrow P_2) \rightarrow T_M}
    \infer[no rule]1{
        \Gamma \uplus \{\langle \Delta ; x \rangle \twoheadrightarrow_p \langle \Delta_{f'} ; f' \rangle : P_1 \rightarrow P_2 \}
        \vdash \langle \Delta_{f'} ; M \rangle \twoheadrightarrow_p \langle \Delta_M ; M' \rangle : T_M
    }
    \infer1[R-App-Compile-Func]{
        \Gamma \vdash
        \langle \Delta_i ; F N \rangle
        \twoheadrightarrow_p
        \langle \Delta_M , \Delta_f , \Delta_{f [k]}; f \, [k] \ f' \rangle : T_M
    }
\end{prooftree}
\]

\[
\begin{prooftree}
    \hypo{(\Delta_f \equiv f :: x : P_1 \rightarrow P_2 \ . \ M : T_M)}
    \infer[no rule]1{(\Delta_{f [k]} \equiv f \, [k] \, \triangleright x : P_1 \rightarrow P_2 \ . \ M' : T_M)}
    \infer[no rule]1{
        \Gamma \vdash
        \langle \Delta , \Delta_f ; N \rangle
        \twoheadrightarrow_p
        \langle \Delta_{f'} , \Delta_f , \Delta_{f [k]} ; f' \rangle : P_1 \rightarrow P_2
    }
    \infer[no rule]1{\Gamma \vdash \langle \Delta_i ; F \rangle \twoheadrightarrow_p \langle \Delta , \Delta_f ; f \rangle : (P_1 \rightarrow P_2) \rightarrow T_M}
    \infer[no rule]1{\Gamma \uplus \{\langle \Delta ; x \rangle \twoheadrightarrow_p \langle \Delta_{f'} ; f' \rangle : P_1 \rightarrow P_2 \}
    \vdash \langle \Delta_{f'} ; M \rangle \twoheadrightarrow_p \langle \Delta_M ; M' \rangle : T_M}
    \infer1[R-App-Ready-Func]{
        \Gamma \vdash
        \langle \Delta_i ; F N \rangle
        \twoheadrightarrow_p
        \langle \Delta_M , \Delta_f , \Delta_{f [k]} ; f \, [k] \ f' \rangle : T_M
    }
\end{prooftree}
\]

\[
\begin{prooftree}
    \hypo{(\Delta_{f [k]} \equiv f \, [k] \triangleright x : T_1 \ . \ M : T_2)}
    \hypo{\Gamma \vdash \langle \Delta , \Delta_{f [k]} ; N \rangle \twoheadrightarrow_p \langle \Delta , \Delta_{f [k]} ; N' \rangle : T_1}
    \infer2[R-App-Compiled]{
        \Gamma \vdash
        \langle \Delta , \Delta_{f [k]} ; f \, [k] \ N \rangle
        \twoheadrightarrow_p
        \langle \Delta , \Delta_{f [k]} ; f \, [k] \ N' \rangle : T_2
    }
\end{prooftree}
\]

The rules \textit{R-App-Compile-*} compile raw functions,
i.e, monomorphize them according to the passed argument.
The rules \textit{R-App-Ready-*} are applied when an appropriate monomorphization is already present in the context.
When one is present, there is no need to create an extra one so that the resulting call references the found monomorphization.

The rules above cover four different types of applications.
The first is when a function is applied with a value of a propertied type that is not a function.
It propagates the propertied value to the raw body of the function and replaces its underlying value by the name of the argument
since the underlying value is received at runtime.
In this case, the function application is transformed into the application of the monomorphized function
and the value of a propertied type is replaced with its underlying value.
The second is when a function is applied with a value of a propertied type that is a function.
In our system, functions are just names that refer to entries in a functional context,
so because it is required to know the name of the function at compile-time to transform the function application,
the underlying value cannot be received at runtime.
The third one is when a function is applied to a function.
For the same reason as in the previous rule,
the name of the function in the argument's position is propagated to the compiling function.
The last one is when the function application was already transformed into a monomorphized version so that it is just transformed into itself.

As the argument of a propertied type in function application is transformed into its underlying value,
the expression \textit{let}, for propertied types to be fully erased at runtime,
must do the same if it gets an expression of a propertied type.

\[
\begin{prooftree}
    \hypo{(T \neq P_1 \rightarrow P_2)}
    \infer[no rule]1{\Gamma \vdash \langle \Delta ; N \rangle \twoheadrightarrow_p \langle \Delta_N ; \text{propertied}[y] \rangle : [T] \langle props \rangle}
    \infer[no rule]1{
        \Gamma \uplus \{ \langle \Delta ; x \rangle \twoheadrightarrow_p \langle \Delta_N ; \text{propertied}[x] \rangle : [T] \langle props \rangle \}
        \vdash \langle \Delta_N ; M \rangle \twoheadrightarrow_p \langle \Delta_M ; M' \rangle : T_M
    }
    \infer1[R-Let-Prop-1]{
        \Gamma \vdash \langle \Delta ; \text{let} \ x = N \ \text{in} \ M \rangle \twoheadrightarrow_p \langle \Delta_M ; \text{let} \ x = y \ \text{in} \ M' \rangle : T_M
    }
\end{prooftree}
\]

\[
\begin{prooftree}
    \hypo{\Gamma \vdash \langle \Delta ; N \rangle \twoheadrightarrow_p \langle \Delta_N ; \text{propertied}[f] \rangle : [P_1 \rightarrow P_2] \langle props \rangle}
    \infer[no rule]1{
        \Gamma \uplus \{ \langle \Delta ; x \rangle \twoheadrightarrow_p \langle \Delta_N ; \text{propertied}[f] \rangle : [P_1 \rightarrow P_2] \langle props \rangle \}
        \vdash \langle \Delta_N ; M \rangle \twoheadrightarrow_p \langle \Delta_M ; M' \rangle : T_M
    }
    \infer1[R-Let-Prop-2]{\Gamma \vdash \langle \Delta ; \text{let} \ x = N \ \text{in} \ M \rangle \twoheadrightarrow_p \langle \Delta_M ; M' \rangle : T_M}
\end{prooftree}
\]

\[
\begin{prooftree}
    \hypo{\Gamma \vdash \langle \Delta ; F \rangle \twoheadrightarrow_p \langle \Delta_F ; f \rangle : T_1 \rightarrow T_2}
    \infer[no rule]1{
        \Gamma \uplus \{\langle \Delta ; x \rangle \twoheadrightarrow_p \langle \Delta_F ; f \rangle : T_1 \rightarrow T_2 \}
            \vdash \langle \Delta_F ; M \rangle \twoheadrightarrow_p \langle \Delta_M ; M' \rangle : T_M
    }
    \infer1[R-Let-Func]{\Gamma \vdash \langle \Delta ; \text{let} \ x = F \ \text{in} \ M \rangle \twoheadrightarrow_p \langle \Delta_M ; M' \rangle : T_M}
\end{prooftree}
\]

The rest of the rules are those which don't transform/evaluate their expressions
but rather propagate the transformation of their subexpressions:

\[
\begin{prooftree}
    \hypo{\Gamma \vdash \langle \Delta ; e_1 \rangle \twoheadrightarrow_p \langle \Delta_{e_1} ; e_1' \rangle : \text{int}}
    \hypo{\Gamma \vdash \langle \Delta_{e_1} ; e_2 \rangle \twoheadrightarrow_p \langle \Delta_{e_2} ; e_2' \rangle : \text{int}}
    \infer2[R-P-Plus]{\Gamma \vdash \langle \Delta ; e_1 + e_2 \rangle \twoheadrightarrow_p \langle \Delta_{e_2} ; e_1' + e_2' \rangle : \text{int}}
\end{prooftree}
\]

\[
\begin{prooftree}
    \hypo{\Gamma \vdash \langle \Delta ; e_1 \rangle \twoheadrightarrow_p \langle \Delta_{e_1} ; e_1' \rangle : \text{int}}
    \hypo{\Gamma \vdash \langle \Delta_{e_1} ; e_2 \rangle \twoheadrightarrow_p \langle \Delta_{e_2} ; e_2' \rangle : \text{int}}
    \infer2[R-P-Minus]{\Gamma \vdash \langle \Delta ; e_1 - e_2 \rangle \twoheadrightarrow_p \langle \Delta_{e_2} ; e_1' - e_2' \rangle : \text{int}}
\end{prooftree}
\]

\[
\begin{prooftree}
    \hypo{(T \neq [T'] \langle \cdots \rangle) \wedge (T \neq P_1 \rightarrow P_2 )}
    \infer[no rule]1{\Gamma \vdash \langle \Delta ; N \rangle \twoheadrightarrow_p \langle \Delta_N ; N' \rangle : T}
    \infer[no rule]1{
        \Gamma \uplus \{ \langle ; x \rangle \twoheadrightarrow_p \langle ; x \rangle : T \}
        \vdash \langle \Delta_N ; M \rangle \twoheadrightarrow_p \langle \Delta_M ; M' \rangle : T_M
    }
    \infer1[R-P-Let]{
        \Gamma \vdash
        \langle \Delta ; \text{let} \ x = N \ \text{in} \ M \rangle
        \twoheadrightarrow_p
        \langle \Delta_M ; \text{let} \ x = N' \ \text{in} \ M' \rangle : T_M
    }
\end{prooftree}
\]

\[
\begin{prooftree}
    \hypo{(f \, [k] \notin \Delta)}
    \infer[no rule]1{(T_1 \neq [T] \langle \cdots \rangle) \wedge (T_1 \neq P_1 \rightarrow P_2)}
    \infer[no rule]1{(\Delta_{f [k]} \equiv f \, [k] \, \triangleright x : T_1 \ . \ M' : T_2)}
    \infer[no rule]1{(\Delta_f \equiv f :: x : T_1 \ . \ M : T_2)}
    \hypo{\Gamma \vdash \langle \Delta_i ; F \rangle \twoheadrightarrow_p \langle \Delta , \Delta_f ; f \rangle : T_1 \rightarrow T_2}
    \infer[no rule]1{
        \Gamma \vdash
        \langle \Delta_M , \Delta_f , \Delta_{f [k]} ; N \rangle
        \twoheadrightarrow_p
        \langle \Delta' ; N' \rangle : T_1
    }
    \infer[no rule]1{
        \Gamma \uplus \{ \langle ; x \rangle \twoheadrightarrow_p \langle ; x \rangle : T_1 \}
        \vdash \langle \Delta ; M \rangle \twoheadrightarrow_p \langle \Delta_M ; M' \rangle : T_2
    }
    \infer2[R-App-Compile]{
        \Gamma \vdash
        \langle \Delta_i ; F N \rangle
        \twoheadrightarrow_p
        \langle \Delta' ; f \, [k] \, \ N' \rangle : T_2
    }
\end{prooftree}
\]

\[
\begin{prooftree}
    \hypo{(T_1 \neq [T] \langle \cdots \rangle) \wedge (T_1 \neq P_1 \rightarrow P_2)}
    \infer[no rule]1{(\Delta_{f [k]} \equiv f \, [k] \, \triangleright x : T_1 \ . \ M' : T_2)}
    \infer[no rule]1{(\Delta_f \equiv f :: x : T_1 \ . \ M : T_2)}
    \infer[no rule]1{
        \Gamma \vdash
        \langle \Delta_M , \Delta_f, \Delta_{f [k]} ; N \rangle
        \twoheadrightarrow_p
        \langle \Delta' ; N' \rangle : T_1
    }
    \infer[no rule]1{\Gamma \vdash \langle \Delta_i ; F \rangle \twoheadrightarrow_p \langle \Delta , \Delta_f, \Delta_{f [k]} ; f \rangle : T_1 \rightarrow T_2}
    \infer[no rule]1{
        \Gamma \uplus \{ \langle ; x \rangle \twoheadrightarrow_p \langle ; x \rangle : T_1 \}
        \vdash \langle \Delta ; M \rangle \twoheadrightarrow_p \langle  \Delta_M ; M' \rangle : T_2
    }
    \infer1[R-App-Ready]{
        \Gamma \vdash
        \langle \Delta_i ; F N \rangle
        \twoheadrightarrow_p
        \langle \Delta' ; f \, [k] \, \ N' \rangle : T_2
    }
\end{prooftree}
\]

And, finally, the rules that make ordinary expressions that work on base data types work on propertied ones:

\[
\begin{prooftree}
    \hypo{\Gamma \vdash \langle \Delta ; e_1 \rangle \twoheadrightarrow_p \langle \Delta_{e_1} ; \text{propertied}[L_1] \rangle : [\text{int}] \langle \cdots \rangle}
    \infer[no rule]1{\Gamma \vdash \langle \Delta_{e_1} ; e_2 \rangle \twoheadrightarrow_p \langle \Delta_{e_2} ; \text{propertied}[L_2] \rangle : [\text{int}] \langle \cdots \rangle}
    \infer1[R-P-Plus-1]{\Gamma \vdash \langle \Delta ; e_1 + e_2 \rangle \twoheadrightarrow_p \langle \Delta_{e_2} ; L_1 + L_2 \rangle : \text{int}}
\end{prooftree}
\]

\[
\begin{prooftree}
    \hypo{\Gamma \vdash \langle \Delta ; e_1 \rangle \twoheadrightarrow_p \langle \Delta_{e_1} ; \text{propertied}[L_1] \rangle : [\text{int}] \langle \cdots \rangle}
    \hypo{\Gamma \vdash \langle \Delta_{e_1} ; e_2 \rangle \twoheadrightarrow_p \langle \Delta_{e_2} ; e_2' \rangle : \text{int}}
    \infer2[R-P-Plus-2]{\Gamma \vdash \langle \Delta ; e_1 + e_2 \rangle \twoheadrightarrow_p \langle \Delta_{e_2} ; L_1 + e_2' \rangle : \text{int}}
\end{prooftree}
\]

\[
\begin{prooftree}
    \hypo{\Gamma \vdash \langle \Delta ; e_1 \rangle \twoheadrightarrow_p \langle \Delta_{e_1} ; e_1' \rangle : \text{int}}
    \hypo{\Gamma \vdash \langle \Delta_{e_1} ; e_2 \rangle \twoheadrightarrow_p \langle \Delta_{e_2} ; \text{propertied}[L_2] \rangle : [\text{int}] \langle \cdots \rangle}
    \infer2[R-P-Plus-3]{\Gamma \vdash \langle \Delta ; e_1 + e_2 \rangle \twoheadrightarrow_p \langle \Delta_{e_2} ; e_1' + L_2 \rangle : \text{int}}
\end{prooftree}
\]

\[
\begin{prooftree}
    \hypo{\Gamma \vdash \langle \Delta ; e_1 \rangle \twoheadrightarrow_p \langle \Delta_{e_1} ; \text{propertied}[L_1] \rangle : [\text{int}] \langle \cdots \rangle}
    \infer[no rule]1{\Gamma \vdash \langle \Delta_{e_1} ; e_2 \rangle \twoheadrightarrow_p \langle \Delta_{e_2} ; \text{propertied}[L_2] \rangle : [\text{int}] \langle \cdots \rangle}
    \infer1[R-P-Minus-1]{\Gamma \vdash \langle \Delta ; e_1 - e_2 \rangle \twoheadrightarrow_p \langle \Delta_{e_2} ; L_1 - L_2 \rangle : \text{int}}
\end{prooftree}
\]

\[
\begin{prooftree}
    \hypo{\Gamma \vdash \langle \Delta ; e_1 \rangle \twoheadrightarrow_p \langle \Delta_{e_1} ; \text{propertied}[L_1] \rangle : [\text{int}] \langle \cdots \rangle}
    \hypo{\Gamma \vdash \langle \Delta_{e_1} ; e_2 \rangle \twoheadrightarrow_p \langle \Delta_{e_2} ; e_2' \rangle : \text{int}}
    \infer2[R-P-Minus-2]{\Gamma \vdash \langle \Delta ; e_1 - e_2 \rangle \twoheadrightarrow_p \langle \Delta_{e_2} ; L_1 - e_2' \rangle : \text{int}}
\end{prooftree}
\]

\[
\begin{prooftree}
    \hypo{\Gamma \vdash \langle \Delta ; e_1 \rangle \twoheadrightarrow_p \langle \Delta_{e_1} ; e_1' \rangle : \text{int}}
    \hypo{\Gamma \vdash \langle \Delta_{e_1} ; e_2 \rangle \twoheadrightarrow_p \langle \Delta_{e_2} ; \text{propertied}[L_2] \rangle : [\text{int}] \langle \cdots \rangle}
    \infer2[R-P-Minus-3]{\Gamma \vdash \langle \Delta ; e_1 - e_2 \rangle \twoheadrightarrow_p \langle \Delta_{e_2} ; e_1' - L_2 \rangle : \text{int}}
\end{prooftree}
\]

\subsection{Operational Semantics}

Now it's time to use the judgment $\twoheadrightarrow_p$ and give a semantics for
expressions that are intended to be evaluated at runtime.
Just as in the previous section, we need to save the bodies of functions somewhere.
For this purpose, we introduce a new context, the context $\Phi$:

\[\Phi ::= ⋄ \ | \ \Phi , f \, [n] \, :: x \otimes M\ \]

Since everything is type checked at compile-time, there is no need to track type information about their bodies and arguments.

In order to ensure that the real program evaluation can be started \textit{only} after
stuff with type properties has been carried out at compile-time,
we give this $\Phi$ context to programs only after a successful transformation and type check:

\[
\begin{prooftree}
    \hypo{\vdash t_1 : T}
    \hypo{\vdash \langle ; t_1 \rangle \twoheadrightarrow_p \langle \Delta ; t_2 \rangle : T}
    \hypo{(T \neq [T'] \langle \cdots \rangle)}
    \hypo{\Phi \equiv \{ (f \, [n] \, \triangleright x : T_1 \ . \ M : T_2) \in \Delta \ | \ f \, [n] \, :: x \otimes M \}}
    \infer4[Ready]{\Phi \triangleright \langle ; t_2 \rangle}
\end{prooftree}
\]

The rule above states that if, under empty typing context,
the program $t_1$ is well-typed, transformed into $t_2$ and $t_2$ is not a value of a propertied type,
then the program $t_2$ is ready to be executed and is given a $\Phi$-context.
We call the program $t_1$ \textit{well-transformed} if it satisfies the conditions above.
The restriction with $t_2$ not being a value of a propertied type is required
because we prohibit all that stuff with type properties to occur in runtime.
In the next section, we will prove that this is the only source of them.

A context that will keep track of variables and their corresponding values is defined as follows:

\[ \sigma ::= ⋄ \ | \ \sigma , x \hookrightarrow v \]

The structural rules for the $\Phi$ context are:

\[
\begin{prooftree}
    \hypo{\Phi \triangleright \langle \sigma ; e \rangle \longmapsto \langle \sigma' ; e' \rangle}
    \hypo{(f \, [n] \notin \Phi)}
    \infer2[Phi-Weak-1]{\Phi , f \, [n] \, :: x \otimes M \triangleright \langle \sigma ; e \rangle \longmapsto \langle \sigma' ; e' \rangle}
\end{prooftree}
\]

\[
\begin{prooftree}
    \hypo{\Phi \triangleright e \ val}
    \hypo{(f \, [n] \notin \Phi)}
    \infer2[Phi-Weak-2]{\Phi , f \, [n] \, :: x \otimes M \triangleright e \ val}
\end{prooftree}
\]

\[
\begin{prooftree}
    \hypo{\Phi \triangleright \langle \sigma ; e \rangle \longmapsto \langle \sigma' ; e' \rangle}
    \hypo{\Phi \triangleright v \ val}
    \hypo{(x \hookrightarrow v \notin \sigma \cup \sigma')}
    \infer3[Sigma-Weak]{\Phi \triangleright \langle \sigma , x \hookrightarrow v ; e \rangle \longmapsto \langle \sigma' , x \hookrightarrow v ; e' \rangle}
\end{prooftree}
\]

\[
\begin{prooftree}
    \hypo{\Phi , \Phi' \triangleright \langle \sigma ; e \rangle \longmapsto \langle \sigma' ; e' \rangle}
    \infer1[Phi-Exchange-1]{\Phi' , \Phi \triangleright \langle \sigma ; e \rangle \longmapsto \langle \sigma' ; e' \rangle}
\end{prooftree}
\]

\[
\begin{prooftree}
    \hypo{\Phi , \Phi' \triangleright e \ val}
    \infer1[Phi-Exchange-2]{\Phi' , \Phi \triangleright e \ val}
\end{prooftree}
\]

\[
\begin{prooftree}
    \hypo{\Phi \triangleright \langle \sigma_1 , \sigma_2 ; e \rangle \longmapsto \langle \sigma_1' , \sigma_2' ; e' \rangle}
    \infer1[Sigma-Exchange]{\Phi \triangleright \langle \sigma_2 , \sigma_1 ; e \rangle \longmapsto \langle \sigma_1' , \sigma_2' ; e' \rangle}
\end{prooftree}
\]

Obtaining a $\Phi$ context for an expression ensures that the expression is ready to process its computation,
so every transition rule implicitly assumes that its source has one.
Since the resulting expression, in order to proceed with the computation, must have the one too,
we added a rule that does just that:

\[
\begin{prooftree}
    \hypo{\Phi \triangleright \langle \sigma ; e \rangle}
    \hypo{\Phi \triangleright \langle \sigma ; e \rangle \longmapsto \langle \sigma' ; e' \rangle}
    \infer2[Phi-Preserve]{\Phi \triangleright \langle \sigma' ; e' \rangle}
\end{prooftree}
\]

We also add two more expressions to the language,
which are not present in the language itself but are responsible
for dropping or retrieving previous values of variables:

\[ e ::= \cdots \ | \ \text{drop} \ x \ \text{after} \ e \ | \ \text{retrieve} \ x = v \ \text{after} \ e \]

Closed values are evaluated by the following rules:

\[
\begin{prooftree}
    \hypo{}
    \infer1[V-Unit]{\triangleright \, () \ val}
\end{prooftree}
\]

\[
\begin{prooftree}
    \hypo{}
    \infer1[V-Int]{\triangleright \, n \ val}
\end{prooftree}
\]

\[
\begin{prooftree}
    \hypo{}
    \infer1[V-Func]{ f \, [n] \, :: x \otimes M \triangleright f \,[n] \ val}
\end{prooftree}
\]

And, finally, the transition rules are:

\[
\begin{prooftree}
    \hypo{}
    \infer1[Var]{\Phi \triangleright \langle x \hookrightarrow v ; x \rangle \longmapsto \langle x \hookrightarrow v ; v \rangle}
\end{prooftree}
\]

\[
\begin{prooftree}
    \hypo{\Phi \triangleright \langle \sigma ; e_1 \rangle \longmapsto \langle \sigma' ; e_1' \rangle}
    \infer1[App-P-1]{\Phi \triangleright \langle \sigma ; e_1 \, e_2 \rangle \longmapsto \langle \sigma' ; e_1' \, e_2 \rangle}
\end{prooftree}
\]

\[
\begin{prooftree}
    \hypo{\Phi \triangleright e_1 \ val}
    \hypo{\Phi \triangleright \langle \sigma ; e_2 \rangle \longmapsto \langle \sigma' ; e_2' \rangle}
    \infer2[App-P-2]{\Phi \triangleright \langle \sigma ; e_1 \, e_2 \rangle \longmapsto \langle \sigma' ; e_1 \, e_2' \rangle}
\end{prooftree}
\]

\[
\begin{prooftree}
    \hypo{\Phi , f \, [n] \, :: x \otimes M \triangleright v \ val}
    \hypo{(x \notin \sigma)}
    \infer2[App-1]{
        \Phi , f \, [n] \, :: x \otimes M \triangleright
        \langle \sigma ; f \, [n] \, \ v \rangle
        \longmapsto
        \langle \sigma , x \hookrightarrow v ; \text{drop} \ x \ \text{after} \ M \rangle
    }
\end{prooftree}
\]

\[
\begin{prooftree}
    \hypo{\Phi , f \, [n] \, :: x \otimes M \triangleright v \ val}
    \infer1[App-2]{
        \Phi , f \, [n] \, :: x \otimes M \triangleright
        \langle \sigma , x \hookrightarrow v_{\text{prev}} ; f \, [n] \, \ v \rangle
        \longmapsto
        \langle \sigma , x \hookrightarrow v ; \text{retrieve} \ x = v_{\text{prev}} \ \text{after} \ M \rangle
    }
\end{prooftree}
\]

\[
\begin{prooftree}
    \infer0[App-With-Func]{
        \Phi , g \, [k] \, :: y \otimes M_g , f \, [n] \, :: x \otimes M_f \triangleright
        \langle \sigma ; f \, [n] \, \ g \rangle
        \longmapsto
        \langle \sigma ; M_f \rangle
    }
\end{prooftree}
\]

\[
\begin{prooftree}
    \hypo{\Phi \triangleright \langle \sigma ; n_1 \rangle \longmapsto \langle \sigma' ; n_1' \rangle}
    \infer1[Plus-P-1]{\Phi \triangleright \langle \sigma ; n_1 + n_2 \rangle \longmapsto \langle \sigma' ; n_1' + n_2 \rangle}
\end{prooftree}
\]

\[
\begin{prooftree}
    \hypo{\Phi \triangleright n_1 \ val}
    \hypo{\Phi \triangleright \langle \sigma ; n_2 \rangle \longmapsto \langle \sigma' ; n_2' \rangle}
    \infer2[Plus-P-2]{\Phi \triangleright \langle \sigma ; n_1 + n_2 \rangle \longmapsto \langle \sigma' ; n_1 + n_2' \rangle}
\end{prooftree}
\]

\[
\begin{prooftree}
    \hypo{\Phi \triangleright n_1 \ val}
    \hypo{\Phi \triangleright n_2 \ val}
    \hypo{(n_1 + n_2 = n)}
    \infer3[Plus]{\Phi \triangleright \langle ; n_1 + n_2 \rangle \longmapsto \langle ; n \rangle}
\end{prooftree}
\]

\[
\begin{prooftree}
    \hypo{\Phi \triangleright \langle \sigma ; n_1 \rangle \longmapsto \langle \sigma' ; n_1' \rangle}
    \infer1[Minus-P-1]{\Phi \triangleright \langle \sigma ; n_1 - n_2 \rangle \longmapsto \langle \sigma' ; n_1' - n_2 \rangle}
\end{prooftree}
\]

\[
\begin{prooftree}
    \hypo{\Phi \triangleright n_1 \ val}
    \hypo{\Phi \triangleright \langle \sigma ; n_2 \rangle \longmapsto \langle \sigma' ; n_2' \rangle}
    \infer2[Minus-P-2]{\Phi \triangleright \langle \sigma ; n_1 - n_2 \rangle \longmapsto \langle \sigma' ; n_1 - n_2' \rangle}
\end{prooftree}
\]

\[
\begin{prooftree}
    \hypo{\Phi \triangleright n_1 \ val}
    \hypo{\Phi \triangleright n_2 \ val}
    \hypo{(n_1 - n_2 = n)}
    \infer3[Minus]{\Phi \triangleright \langle ; n_1 - n_2 \rangle \longmapsto \langle ; n \rangle}
\end{prooftree}
\]

\[
\begin{prooftree}
    \hypo{\Phi \triangleright \langle \sigma ; e \rangle \longmapsto \langle \sigma' ; e' \rangle}
    \infer1[Let-P]{\Phi \triangleright \langle \sigma ; \text{let} \ x = e \ \text{in} \ M \rangle \longmapsto \langle \sigma' ; \text{let} \ x = e' \ \text{in} \ M \rangle}
\end{prooftree}
\]

\[
\begin{prooftree}
    \hypo{\Phi \triangleright v \ val}
    \hypo{(x \notin \sigma)}
    \infer2[Let-1]{
        \Phi \triangleright
        \langle \sigma ; \text{let} \ x = v \ \text{in} \ M \rangle
        \longmapsto
        \langle \sigma , x \hookrightarrow v ; \text{drop} \ x \ \text{after} \ M \rangle
    }
\end{prooftree}
\]

\[
\begin{prooftree}
    \hypo{\Phi \triangleright v \ val}
    \infer1[Let-2]{
        \Phi \triangleright
        \langle \sigma , x \hookrightarrow v_{\text{prev}}; \text{let} \ x = v \ \text{in} \ M \rangle
        \longmapsto
        \langle \sigma , x \hookrightarrow v ; \text{retrieve} \ x = v_{\text{prev}}\ \text{after} \ M \rangle
    }
\end{prooftree}
\]

\[
\begin{prooftree}
    \hypo{\Phi \triangleright \langle \sigma ; e \rangle \longmapsto \langle \sigma' ; e' \rangle}
    \infer1[Drop-After-1]{
        \Phi \triangleright
        \langle \sigma ; \text{drop} \ x \ \text{after} \ e \rangle
        \longmapsto
        \langle \sigma' ; \text{drop} \ x \ \text{after} \ e' \rangle
    }
\end{prooftree}
\]

\[
\begin{prooftree}
    \hypo{\Phi \triangleright v \ val}
    \infer1[Drop-After-2]{
        \Phi \triangleright
        \langle \sigma , x \hookrightarrow v_x ; \text{drop} \ x \ \text{after} \ v \rangle
        \longmapsto
        \langle \sigma ; v \rangle
    }
\end{prooftree}
\]

\[
\begin{prooftree}
    \hypo{\Phi \triangleright \langle \sigma ; e \rangle \longmapsto \langle \sigma' ; e' \rangle}
    \infer1[Retrieve-After-1]{
        \Phi \triangleright
        \langle \sigma ; \text{retrieve} \ x = v \ \text{after} \ e \rangle
        \longmapsto
        \langle \sigma' ; \text{retrieve} \ x = v \ \text{after} \ e' \rangle
    }
\end{prooftree}
\]

\[
\begin{prooftree}
    \hypo{\Phi \triangleright v \ val}
    \infer1[Retrieve-After-2]{
        \Phi \triangleright
        \langle \sigma , x \hookrightarrow v_x ; \text{retrieve} \ x = v_{\text{prev}} \ \text{after} \ v \rangle
        \longmapsto
        \langle \sigma , x \hookrightarrow v_{\text{prev}} ; v \rangle
    }
\end{prooftree}
\]

\section{Properties of $\lambda_{\twoheadrightarrow_p}$}

In this section, we are going to explore some important properties of $\lambda_{\twoheadrightarrow_p}$.
We start with the properties of compile-time program transformation.

\subsection{Program transformation properties}

One of the important claims that we made in previous sections was that no values
of propertied types and propertied types themselves are present at runtime.
That is, everything related to them must be carried out at compile-time.
We start with the lemma that states just that.

\paragraph{Lemma 4.1}
\textit{
    If $\ \Gamma \vdash t_1 : T$ and $\ \Gamma \vdash \langle \Delta ; t_1 \rangle \twoheadrightarrow_p \langle \Delta' ; t_2 \rangle : T$,
    then either $T$ is of the form or no subexpression in $t_2$ is of a type $[T'] \langle \cdots \rangle$.
    What's more, for every function $f \, [n] \triangleright x : T_1 \ . \ M : T_2$ in $\Delta'$,
    neither $T_1$, $T_2$, nor any subexpression of $M$ is of the form $[T'] \langle \cdots \rangle$.
}

The allowance for $t_2$ be a value of a propertied type is justified by the rule \textit{Ready}
from the previous section, since no $\Phi$ context, which is required for an expression to be passed to runtime,
is given for one.

\paragraph{Proof 4.1}
\textit{By induction on the derivation rules of the judgment $\twoheadrightarrow_p$ and the typing rules from Section 3.2}.

The proof of the fact that neither $T_1$ nor $T_2$ of $f \, [n] \triangleright x : T_1 \ . \ M : T_2$ may
be of the form $[T'] \langle \cdots \rangle$ is immediate by the typing rule I-Func,
which states that in order to form a function, the resulting type must not be a propertied one.
$T_1$ a priori cannot be a propertied type since there is no conventional syntax for them in the language.

For the rules that act on values — \textit{R-V-Unit} and \textit{R-V-Int} —
the proof is immediate since they are transformed into themselves.
The rules that act on the expression \textit{set} make it be of type $[T'] \langle \cdots \rangle$, so the proof is immediate too.
Next, the rules \textit{R-Get} and \textit{R-Ext} transform their expressions into their subexpressions,
which are, by induction, assumed to have the desired property.

The proof that the same holds for every function in the context $\Delta'$ is obtained
by the induction on the rules that perform monomorphization and add them to functional contexts.

What's interesting are the rules \textit{R-App-*-Prop-1} — when the argument
of the function application is of a propertied type,
the expression is transformed into a runtime application
with the argument being the underlying value.
Since that, we got rid of the propertied type in the argument.
The same situation is found with the rules that act on the expression \textit{if-has} —
there are special rules that cover cases when the property being extracted is of a propertied type —
they just transform them to their underlying values.

Proofs for the rest of the rules follow the same structure and are obtained by using the same techniques.
So, in order not to litter this paragraph, we omitted them here.

The next important property of the $\twoheadrightarrow_p$-transformation is that it must be deterministic,
since none of our compile-time constructions is expected to do something that causes non-deterministic behavior.

\paragraph{Lemma 4.2}
\textit{
    If $\ \Gamma \vdash \langle \Delta ; t_1 \rangle \twoheadrightarrow_p \langle \Delta_1 ; t_2 \rangle : T$
    and $\ \Gamma \vdash \langle \Delta ; t_1 \rangle \twoheadrightarrow_p \langle \Delta_2 ; t_2' \rangle : T'$,
    then $\ \Delta_1 = \Delta_2$, $\ t_2 = t_2'$ and $\ T = T'$
}

We didn't define equality judgment for types, terms, and contexts,
so what we mean is the ordinary syntactic one.

\paragraph{Proof 4.2}
\textit{By induction on the derivation rules of the judgment $\twoheadrightarrow_p$}.

The proof has a similar structure to the one of the \textit{Lemma 4.1} —
for some expressions, there is only one rule defining their transformations,
so assuming that all subexpressions are deterministic makes the proof for them.

Other rules explore the structure of their subexpressions so that no expression can be transformed by
two rules simultaneously.
For instance, let's take a look at the rules \textit{R-Let-Prop-1},
\textit{R-Let-Prop-2}, \textit{R-Let-Func}, and \textit{R-P-Let}.
They all operate on an expression $let \; x = N \; in \; M$, but:
\begin{itemize}
    \item \textit{R-Let-Prop-2} works only when \textit{N} is of a propertied type,
          and the underlying value is of a function one.
    \item \textit{R-Let-Prop-1} works only when \textit{N} is of a propertied type too,
          but the underlying value is \textit{not} of a function type, thus covering
          all cases that \textit{R-Let-Prop-2} excludes.
    \item \textit{R-Let-Func} works only when \textit{N} is transformed into a function.
    \item \textit{R-P-Let} works only when \textit{N} is neither transformed into a function nor to a value of a propertied type.
\end{itemize}
thus mutually excluding each other.

\subsection{Type soundness}

Now, when it was shown that the compile-time part has the properties we needed,
we are able to prove one of the most important properties of a programming language —
the property of being \textit{type sound}.
The \textit{type soundness} theorem states that if a program passes its programming language's
type checker, then it is guaranteed that it has a well-defined behavior when executed.
It consists of two theorems — the first, called the \textit{progress} theorem,
which states that if an expression $e$ is well-typed
(and, in our case, well-transformed under $\twoheadrightarrow_p$)
then either it is already a value, or we can proceed with its computation.
And the second, called the \textit{preservation} theorem, which states that
if an expression $e$ has the type $T$ and $\Phi \triangleright \langle \sigma ; e \rangle \longmapsto \langle \sigma' ; e' \rangle$, then $e'$ is of type $T$ too.

\paragraph{Lemma 4.3 (Progress)}
\textit{
    If $\ \Phi \triangleright \langle \sigma ; e \rangle$,
    then either $\ \Phi \triangleright \langle \sigma ; e \rangle \longmapsto \langle \sigma' ; e' \rangle\ $
    or $\ \Phi \triangleright e \; val$
}

\paragraph{Proof 4.3}
\textit{By induction on the rules given in Section 3.4}

The proof for expressions that represent values is immediate.
Next, one must observe that the semantics is defined for all expressions
except ones that carry out stuff with type properties, such as \textit{set}, \textit{get}, etc.
But since we proved (the lemma 4.1) that all stuff related to them is carried out at compile-time,
no expression like \textit{set} and \textit{get} can occur at runtime, so the assumption is justified.

\paragraph{Lemma 4.4 (Preservation)}
\textit{if $\ \Gamma \vdash e : T$ and $\ \Phi \triangleright \langle \sigma ; e \rangle \longmapsto \langle \sigma' ; e' \rangle$, then $\Gamma \vdash e' : T$}

\paragraph{Proof 4.4}
\textit{By induction on the typing rules from Section 3.2 and the rules given in Section 3.4}

\section{Programming in $\lambda_{\twoheadrightarrow_p}$}

In this section, we are going to write two trivial programs in $\lambda_{\twoheadrightarrow_p}$
and prove that they evaluate to some specific expressions.
This section intends to make the processing of type properties more clear,
by performing a step-by-step type checking, transformation,
and runtime evaluation of the programs.
The reader is free to skip this section.

We start with a very simple program to make the reader familiar with the structure of the proofs.

\paragraph{Lemma 5.1}
\textit{Let $L$ be the program}
\begin{lstlisting}
  let y = 5 in
  func f x : int with
    x + y in
  f 1
\end{lstlisting}

\textit{$L$ evaluates to 6.}

Formally, if a program $L$ \textit{evaluates} to an expression $L'$, then it means that
$L$ is well-typed, well-transformed, and $\Phi \triangleright \langle ; L \rangle \longmapsto^* \langle \sigma ; L' \rangle$,
where $\Phi \triangleright \langle ; L \rangle \longmapsto^* \langle \sigma ; L' \rangle$ stands for
$\Phi \triangleright \langle ; L \rangle \longmapsto \langle \sigma_1 ; e_1 \rangle$,
$\Phi \triangleright \langle \sigma_1 ; e_1 \rangle \longmapsto \langle \sigma_2 ; e_2 \rangle$,
...,
$\Phi \triangleright \langle \sigma_n ; e_n \rangle \longmapsto \langle \sigma ; L' \rangle$
with $n \geq 0$, so we reformulate our lemma as follows:

\textit{
    Given the program $L$, $L$ is well-typed, well-transformed,
    and $\Phi \triangleright \langle ; L \rangle \longmapsto^* \langle \sigma ; 6 \rangle$
}.

\paragraph{Proof 5.1}

We first start with proof that the program is well-typed.
Basically, what we want is to derive the judgment

\[\vdash \text{let} \ y = 5 \ \text{in} \ \text{func} \ f \ x : \text{int} \ \text{with} \ x + y \ \text{in} \ f \, 1 : \text{int}\]

According to the rule \textit{E-Let}, this can be derived if we have the following judgments entailed:
\[y : \text{int} \vdash \text{func} \ f \ x : \text{int} \ \text{with} \ x + y \ \text{in} \ f \, 1 : \text{int}\]
\[\vdash 5 : \text{int}\]

The second rule is derived immediately by the rule \textit{I-Int}.
The first, according to the rule \textit{I-Func}, is obtained when we have that
\[y : \text{int} , x : \text{int} \vdash x + y : \text{int}\]
\[y : \text{int} , x : [\text{int}] \langle \rangle \vdash x + y : \text{int}\]
\[y : \text{int} , f : \text{int} \rightarrow \text{int} \vdash f \, 1 : \text{int}\]

which are all easily obtained by \textit{E-Plus}, \textit{E-App-1}, and some structural rules
so that the proof is completed.

What we need to prove next is that this program is well-transformed,
which is expressed by obtaining the judgment
\[\vdash \langle ; L \rangle \twoheadrightarrow_p \langle \Delta ; t_2 \rangle : \text{int}\]

By the rule \textit{R-P-Let}, if the type of $y$ is not a propertied or function one
(which is exactly the case here, since $y$ is of type int),
$\langle ; \text{let} \ y = 5 \ \text{in} \ M \rangle$ is transformed into
$\langle \Delta ; \text{let} \ y = N' \ \text{in} \ M' \rangle$, where $N'$, $M'$, and $\Delta$ are:
\[\vdash \langle ; 5 \rangle \twoheadrightarrow_p \langle \Delta_1 ; N' \rangle : \text{int}\]
\[
  \langle ; y \rangle \twoheadrightarrow_p \langle ; y \rangle : \text{int}
  \vdash
  \langle \Delta_1 ; \text{func} \ f \ x : \text{int} \ \text{with} \ x + y \ \text{in} \ f \, 1 \rangle
  \twoheadrightarrow_p
  \langle \Delta ; M' \rangle : \text{int}
\]

For the first judgment, by the rule \textit{R-P-Int} and the fact that
transformations under $\twoheadrightarrow_p$ are deterministic
(which was proven in the previous section),
it must be the case that $N'$ is $5$.

The second is obtained by the rule \textit{R-Func}, which requires us to show that

\[ \langle ; y \rangle \twoheadrightarrow_p \langle ; y \rangle : \text{int} , x : \text{int} \vdash x + y : \text{int}\]
\[ \langle ; y \rangle \twoheadrightarrow_p \langle ; y \rangle : \text{int} , f : \text{int} \rightarrow \text{int} \vdash
    \langle f :: x : \text{int} \ . \ x + y : \text{int} ; f \, 1 \rangle \twoheadrightarrow_p \langle \Delta ; M' \rangle : \text{int} \]

The first is obtained easily by the rules \textit{R-S-Var} and \textit{E-Plus}.
Let:

\[\Gamma \equiv \langle ; y \rangle \twoheadrightarrow_p \langle ; y \rangle : \text{int} , f : \text{int} \rightarrow \text{int},\]
\[\Delta_f \equiv f :: x : \text{int} \ . \ x + y : \text{int},\]
\[\Delta_{f [1]} \equiv f \, [1] \triangleright x : \text{int} \ . \ x + y : \text{int}\]

Then, the required premiss is easily obtained by the rule \textit{R-App-Compile}:

\[
\begin{prooftree}
    \hypo{\Gamma \vdash \langle \Delta_f ; f \rangle \twoheadrightarrow_p \langle \Delta_f ; f \rangle : \text{int} \rightarrow \text{int}}
    \infer[no rule]1{\Gamma \vdash \langle \Delta_f , \Delta_{f [1]}; 1 \rangle \twoheadrightarrow_p \langle \Delta_f , \Delta_{f [1]}; 1 \rangle : \text{int}}
    \infer[no rule]1{
        \Gamma \uplus \{ \langle ; x \rangle \twoheadrightarrow_p \langle ; x \rangle : \text{int} \}
        \vdash \langle ; x + y \rangle \twoheadrightarrow_p \langle ; x + y \rangle : \text{int}
    }
    \infer1{\Gamma \vdash \langle \Delta_f ; f \, 1 \rangle \twoheadrightarrow_p \langle \Delta_f , \Delta_{f [1]}; f \, [1] \ 1 \rangle : \text{int}}
\end{prooftree}
\]

As the required premises are obtained easily, the proof is completed.
The final transformation judgment is:

\[
\vdash \langle ; \text{let} \ y = 5 \ \text{in} \ \text{func} \ f \ x : \text{int} \ \text{with} \ x + y \ \text{in} \ f \, 1 \rangle
\twoheadrightarrow_p
\langle \Delta_f , \Delta_{f [1]} ; \text{let} \ y = 5 \ \text{in} \ f \, [1] \ 1 \rangle : \text{int}
\]

Since $int$ is not a propertied type, this completes the proof and justifies the assumption that the program is well-transformed.

Now, all is left is to prove that, under the operational semantics we gave in section 3.4,
this program is evaluated to the term 6.
First, we need to obtain a $\Phi$ context for our program, which is,
by the rule \textit{Ready}, easily obtained under the assumption that our program is well-typed and well-transformed:

\[
\begin{prooftree}
    \hypo{\vdash L : \text{int}}
    \hypo{\vdash \langle ; L \rangle \twoheadrightarrow_p \langle \Delta ; t_2 \rangle : \text{int}}
    \infer2[Ready]{\Phi \triangleright \langle ; t_2 \rangle}
\end{prooftree}
\]

where $\Phi$, in our case, is just $f \, [1]  :: x \otimes x + y$.

When the ready context is obtained, the evaluation of the program proceeds as follows:

\[
\begin{prooftree}
    \infer0{\Phi \triangleright 5 \ val}
    \infer1{
        \Phi \triangleright
            \langle ; \text{let} \ y = 5 \ \text{in} \ f \, [1] \ 1 \rangle
            \longmapsto
            \langle y \hookrightarrow 5 ; \text{drop} \ y \ \text{after} \ f \, [1] \ 1 \rangle
    }
    \infer0{\Phi \triangleright 1 \ val}
    \infer2{
        f \, [1] :: x \otimes x + y \triangleright
            \langle y \hookrightarrow 5 ; f \, [1] \ 1 \rangle
            \longmapsto
            \langle y \hookrightarrow 5 , x \hookrightarrow 1 ;
                    \text{drop} \ x \ \text{after} \ x + y \rangle
    }
    \infer1{\sigma \equiv y \hookrightarrow 5 , x \hookrightarrow 1}
    \infer1{\Phi \triangleright \langle \sigma ; x \rangle \longmapsto \langle \sigma ; 1 \rangle}
    \infer1{\Phi \triangleright \langle \sigma ; x + y \rangle \longmapsto \langle \sigma ; 1 + y \rangle}
    \infer1{\Phi \triangleright \langle \sigma ; y \rangle \longmapsto \langle \sigma ; 5 \rangle}
    \infer1{\Phi \triangleright \langle \sigma ; 1 + y \rangle \longmapsto \langle \sigma ; 1 + 5 \rangle}
    \infer1{\Phi \triangleright \langle ; 1 + 5 \rangle \longmapsto \langle ; 6 \rangle}
    \infer1{\Phi \triangleright \langle \sigma ; 1 + 5 \rangle \longmapsto \langle \sigma ; 6 \rangle}
    \infer1{\Phi \triangleright 6 \ val}
    \infer1{
        \Phi \triangleright
            \langle \sigma ; \text{drop} \ x \ \text{after} \ x + y \rangle
            \longmapsto^*
            \langle \sigma ; \text{drop} \ x \ \text{after} \ 6 \rangle
    }
    \infer1{
        \Phi \triangleright
            \langle y \hookrightarrow 5 , x \hookrightarrow 1 ; \text{drop} \ x \ \text{after} \ 6 \rangle
            \longmapsto
            \langle y \hookrightarrow 5 ; 6 \rangle
    }
    \infer1{
        \Phi \triangleright
            \langle y \hookrightarrow 5 ; \text{drop} \ y \ \text{after} \ f \, [1] \ 1 \rangle
            \longmapsto^*
            \langle y \hookrightarrow 5 ; \text{drop} \ y \ \text{after} \ 6 \rangle
    }
    \infer1{
        \Phi \triangleright
            \langle y \hookrightarrow 5 ; \text{drop} \ y \ \text{after} \ 6 \rangle
            \longmapsto
            \langle ; 6 \rangle
    }
\end{prooftree}
\]

This completes the proof of the \textit{lemma 5.1}.

Now we shall explore more complicated programs.
Let $L$ be the program:
\begin{lstlisting}[frame=tlrb]
  func f x : int with
    if-has x c : int bind-as c in
        c + 1
    else extract(x) in
  let y = set(5, c, 5) in
  f y
\end{lstlisting}

\paragraph{Lemma 5.2}
\textit{$L$ is well-typed, well-transformed, and $\Phi \triangleright \langle ; L \rangle \longmapsto^* \langle \sigma ; 6 \rangle$}.

\paragraph{Proof 5.2}

We left the proof that the program is well-typed to the reader,
since it is pretty trivial and follows the same strategy as the one of the \textit{lemma 5.1}.

What we want to focus on now is the property of the program of being well-transformed —
since the program utilizes type properties, exploring the property will make it clear
how exactly they are carried out at compile-time.
In particular, just as with the \textit{lemma 5.1}, all we want is to derive the judgment

\[\vdash \langle ; L \rangle \twoheadrightarrow_p \langle \Delta ; L' \rangle : T\]

and ensure that $T$ is not a propertied type.

The program $L$ contains 6 lines of code, so repeatedly writing its parts
every time a new assumption arises would be quite unreadable and space-consuming,
so, it would be reasonable to name distinct parts of the program and refer to them when we do our proof.

\[L \equiv \text{func} \ f \ x : \text{int} \ \text{with} \ L_1 \ \text{in} \ L_2\]
\[L_1 = \text{if-has} \ x \, c : \text{int} \ \text{bind-as} \ c \ \text{in} \ c + 1 \ \text{else} \ \text{extract}(x)\]
\[L_2 = \text{let} \ y = \text{set}(5, \, c, \, 5) \ \text{in} \ f \, y\]

The only rule that can be applied to the program $L$ is the rule \textit{R-Func},
which says that to obtain the judgment

\[
\vdash
    \langle ; \text{func} \ f \ x : \text{int} \ \text{with} \ L_1 \ \text{in} \ L_2 \rangle
    \twoheadrightarrow_p
    \langle \Delta ; L_2' \rangle : T,
\]

it is enough to show that

\[x : \text{int} \vdash L_1 : T_2\]
\[x : [\text{int}] \langle \rangle \vdash L_1 : T_2\]
\[f : T_1 \rightarrow T_2 \vdash \langle f :: x : T_1 \ . \ L_1 : T_2 ; L_2 \rangle \twoheadrightarrow_p \langle \Delta ; L_2' \rangle : T \]

The first two are easily obtained by the rules \textit{E-If-Has-1} \textit{E-If-Has-2}.
Let $\Gamma \equiv f : T_1 \rightarrow T_2$ and $\Delta_1 \equiv f :: x : T_1 \ . \ L_1 : T_2$.

To obtain the second, according to the rule \textit{R-Let-Prop-1},
it is enough to show that:

\[
\Gamma \vdash \langle \Delta_1 ; \text{set}(5, \, c, \, 5) \rangle \twoheadrightarrow_p \langle \Delta_1 ; \text{propertied}[5] \rangle
    : [\text{int}] \langle c \hookrightarrow 5[\text{int}] \rangle
\]
\[
\Gamma \uplus \{ \langle \Delta_1 ; y \rangle \twoheadrightarrow_p \langle \Delta_1 ; \text{propertied}[y] \rangle : [\text{int}] \langle c \hookrightarrow 5[\text{int}] \rangle \}
    \vdash \langle \Delta_1 ; f \, y \rangle \twoheadrightarrow_p \langle \Delta ; F' \rangle : T
\]

The first is immediate by the rules \textit{R-Set-1}, \textit{R-V-Int}, and corresponding structural ones.
Let

\[\Gamma' \equiv \Gamma , \langle \Delta_1 ; y \rangle \twoheadrightarrow_p \langle \Delta_1 ; \text{propertied}[y] \rangle : [\text{int}] \langle c \hookrightarrow 5[\text{int}] \rangle\]

Then, the second is easily obtained by the rule \textit{R-App-Compile-Prop-1}:

\[\Gamma' \vdash \langle \Delta_1 ; f \, y \rangle \twoheadrightarrow_p \langle \Delta_1 , f \, [1] \triangleright x : \text{int} \ . \ L_1' : T_2 ; f \, [1] \ y \rangle : T_2\]

which requires us to show that

\[\Gamma' \vdash \langle \Delta_1 ; f \rangle \twoheadrightarrow_p \langle \Delta_1 ; f \rangle : \text{int} \rightarrow T_2\]
\[\Gamma' \vdash \langle \Delta_1 ; y \rangle \twoheadrightarrow_p \langle \Delta_1 ; \text{propertied}[y] \rangle : [\text{int}] \langle c \hookrightarrow 5[\text{int}] \rangle\]
\[
\Gamma' \uplus \{ \langle ; x \rangle \twoheadrightarrow_p \langle ; \text{propertied}[x] \rangle : [\text{int}] \langle c \hookrightarrow 5[\text{int}] \rangle \} \vdash
    \langle ; L_1 \rangle \twoheadrightarrow_p \langle ; L_1' \rangle : T_2
\]

where $T_1$ becomes int.

The first two are trivial.
Let
\[\Gamma'' \equiv \Gamma' , \langle ; x \rangle \twoheadrightarrow_p \langle ; \text{propertied}[x] \rangle: [\text{int}] \langle c \hookrightarrow 5[\text{int}] \rangle\]

The last one, according to the rule \textit{R-If-Has-4}, can be obtained by obtaining:

\[\Gamma'' \vdash \langle ; x \rangle \twoheadrightarrow_p \langle ; \text{propertied}[x] \rangle : [\text{int}] \langle c \hookrightarrow 5[\text{int}] \rangle\]
\[\Gamma'' \vdash \langle ; 5 \rangle \twoheadrightarrow_p \langle ; 5 \rangle : \text{int}\]
\[\Gamma'' , \langle ; c \rangle \twoheadrightarrow_p \langle ; c \rangle : \text{int} \vdash \langle ; c + 1 \rangle \twoheadrightarrow_p \langle ; c + 1 \rangle : \text{int}\]

which are all obtained pretty easily and where $T_2$ become int.
The final transformation judgment is:

\[\vdash \langle ; L \rangle \twoheadrightarrow_p \langle \Delta ; \text{let} \ y = 5 \ \text{in} \ f \, [1] \ y \rangle : \text{int}\]

with $\Delta$ being $f :: x : \text{int} \ . \ L_1 : \text{int} , f \, [1] \triangleright x : \text{int} \ . \ \text{let} \ c = 5 \ \text{in} \ c + 1 : \text{int}$.

Now all is left is to prove that $L'$ evaluates to 6.
Since after transformation, the program became no more complicated than the one from the \text{lemma 5.1}, we left it to the reader.

\end{document}